\long\def\symbolfootnote[#1]#2{\begingroup%
\def\thefootnote{\fnsymbol{footnote}}\footnote[#1]{#2}\endgroup} 
\def\aj{AJ}%          % Astronomical Journal
\def\araa{ARA\&A}%          % Annual Review of Astron and Astrophys
\def\apj{ApJ}%          % Astrophysical Journal
\def\apjs{ApJS}%          % Astrophysical Journal, Supplement
\def\aap{A\&A}%          % Astronomy and Astrophysics
\def\aapr{A\&AR}%          % Astronomy and Astrophysics Reviews
\def\mnras{MNRAS}%          % Monthly Notices of the RAS
\def\na{New Astron.}%          % New Astronomy
\def\pasp{PASP}%          % Publications of the ASP
\def\pasj{PASJ}%          % Publications of the ASJ
\newcommand{\emacss}{{\small{\textsc{EMACSS}}}}
\newcommand{\dr}{{\rm d}}
\newcommand{\erf}{{\rm erf}}

\newcommand{\etide}{E_{\rm tide}}

\newcommand{\F}{\mathcal{F}}
\newcommand{\K}{\mathcal{K}}

\newcommand{\msunpc}{M_\odot\,{\rm pc}^{-3}}

\newcommand{\rc}{r_{\rm c}}

\newcommand{\rcdot}{\dot{r}_{\rm c}}
\newcommand{\rcrh}{\mathcal{R}_{\rm ch}}
\newcommand{\rcrhmin}{\mathcal{R}_{\rm ch}^{\rm min}}
\newcommand{\rcrhdot}{\dot{\mathcal{R}}_{\rm ch}}
\newcommand{\rcrhn}{\mathcal{R}_{\rm ch0}}
\newcommand{\rh}{r_{\rm h}}
\newcommand{\rv}{r_{\rm v}}
\newcommand{\rhdot}{\dot{r}_{\rm h}}
\newcommand{\rhrj}{\mathcal{R}_{\rm hJ}}

\newcommand{\rvrj}{\mathcal{R}_{\rm vJ}}

\newcommand{\rvrjo}{\mathcal{R}_{\rm vJ1}}
\newcommand{\rhoc}{\rho_{\rm c}}
\newcommand{\rhon}{\rho_{\rm 0}}
\newcommand{\rhocn}{\rho_{\rm c0}}
\newcommand{\rhocdot}{\dot{\rho}_{\rm c}}
\newcommand{\rhoh}{\rho_{\rm h}}

\newcommand{\rt}{r_{\rm t}}
\newcommand{\rj}{r_{\rm J}}
\newcommand{\PP}{\mathcal{P}}

\newcommand{\edot}{\dot{E}}

\newcommand{\mc}{M_{\rm c}}

\newcommand{\xie}{\xi_{\rm e}}
\newcommand{\xieo}{\xi_{\rm e1}}
\newcommand{\sigmac}{\sigma_{\rm c}}
\newcommand{\trc}{\tau_{\rm rc}}
\newcommand{\trh}{\tau_{\rm rh}}

\documentclass[useAMS,usenatbib, times]{mn2e}
\usepackage{amssymb,latexsym,graphicx,natbib,eufrak,times,amsmath}

\title[A code for the evolution of star clusters II]
  {A prescription and fast code for the long-term evolution of star clusters -- II.  Unbalanced and core  evolution}
\author[Gieles, Alexander, Lamers \& Baumgardt]
  {Mark~Gieles$^1$, Poul E.R. Alexander $^2$, Henny~J.G.L.M. Lamers$^3$ and Holger Baumgardt$^4$\\
$^1$ Department of Physics, University of Surrey, Guildford, GU2 7XH, UK\\
$^2$ Institute of Astronomy, University of Cambridge, Madingley Road, Cambridge, CB3 0HA, UK\\
 $^3$  Astronomical Institute Anton Pannekoek, University of Amsterdam, PO Box 94249, 1090\,GE Amsterdam, The Netherlands\\
 $^4$ School of Mathematics and Physics, University of Queensland, St. Lucia, QLD 4072, Brisbane, Australia \\
}
\date{Accepted 2013 October 9.  Received 2013 October 8; in original form 2013 September 3}

\def\LaTeX{L\kern-.36em\raise.3ex\hbox{a}\kern-.15em
    T\kern-.1667em\lower.7ex\hbox{E}\kern-.125emX}

\begin{document}         
\maketitle
\begin{abstract}
We  introduce version two of the fast star cluster evolution code Evolve Me A Cluster of StarS (\emacss).  The first version (Alexander and Gieles)  assumed that cluster evolution is balanced for the majority of the life-cycle, meaning that the rate of energy generation in the core of the cluster equals the diffusion rate of energy by two-body relaxation, which makes the code suitable for modelling clusters in  weak tidal fields.  In this new version, we extend the model to include an unbalanced phase of evolution  to describe the pre-collapse evolution and the accompanying  escape rate such that clusters in strong tidal fields can also be modelled. We also add a prescription for the evolution of the core radius and density and a related cluster concentration parameter. The model  simultaneously solves a series of first-order ordinary differential equations for the rate of change of the core radius, half-mass radius and the number of  member stars $N$. About two thousand integration steps in time are required to solve for the entire evolution of a star cluster and this number is approximately independent of $N$. We compare the model to the variation of these parameters following from a series of direct $N$-body calculations of single-mass clusters and find good agreement in the evolution of all parameters. Relevant time-scales, such as the total lifetimes and core collapse times, are reproduced with an accuracy of about 10\% for clusters with various initial half-mass radii (relative to their Jacobi radii) and a range of different initial $N$ up to $N = 65\,536$.  The current version of \emacss\ contains the basic physics that allows us to evolve several cluster properties for single-mass clusters in a simple and fast way. We intend to extend this framework to include more realistic initial conditions, such as a stellar mass spectrum and mass-loss from stars. The \emacss\ code can be used  in star cluster population studies and in models that consider the co-evolution of (globular) star clusters and large scale structures. 
\end{abstract}

%%%%%%%%%%%%%%%%%%%%%%%%%%%%%%%%%%%%%%%%%%%%
\section{Introduction}
The dynamical evolution of star clusters is the result of several internal and external processes, including two-body relaxation, interactions between single and binary stars, escape across the tidal boundary and the internal evolution and mass-loss of single and binary stars \citep[e.g.][]{1997A&ARv...8....1M}. Modelling collisional systems is challenging because all these effects operate on their own time-scale, ranging over many orders of magnitudes from the orbital period of hard binary stars to the Galactic orbit of the cluster, and depending in different ways on the number of stars $N$ \citep{1998MNRAS.297..794A}. The direct $N$-body approach is a versatile method for solving the gravitational $N$-body problem and correctly combines the interplay between the various dynamical scaling laws and their corresponding time-scales. 
Owing to recent progress in the use of special hardware to accelerate the force calculations  \citep{2009NewA...14..630G,2012MNRAS.424..545N} it is now feasible to model medium sized globular clusters ($N\simeq2-3\times10^5$), with moderate initial densities, over a Hubble  \citep[][]{2012MNRAS.425.2872H, 2013MNRAS.430L..30S}. However, 
the $\mathcal{O}(N^2)$ nature of the computational effort of direct $N$-body integrations does not allow us yet to model globular clusters containing the number of stars of typical globular clusters (about $10^6$) with realistic initial density ($\gtrsim10^4\,\msunpc$) over a Hubble time

We aim to develop a relatively simple, and extremely fast (compared to the direct $N$-body approach) prescription for  the evolution of a few fundamental  properties of tidally limited clusters, such as $N$ and the various cluster radii (core radius, half-mass radius and tidal radius) with an $N$-independent computational effort. Having a fast and simplified prescription of complex astrophysical objects allows us to use  these objects in population synthesis studies, or to combine the  evolutionary prescription with that of other astrophysical phenomena. Examples  exist for other applications, for example, for the evolution of individual stars of different mass and  metallicity   \citep[][]{2000MNRAS.315..543H},   binary stars \citep{2002MNRAS.329..897H}, binary populations \citep*{2008MNRAS.384.1109E} and for the products of stellar collisions \citep{2002ApJ...568..939L}. A possible application of such a tool for star cluster evolution  is the modelling of observed properties of star cluster populations \citep{2005ApJ...634.1002J, 2007ApJS..171..101J,2013ApJ...772...82H}, which will enable us to  use  star clusters more efficiently  as tracers of the formation and evolution of the host galaxy \citep*{2002ARA&A..40..487F,2006ARA&A..44..193B, 2008ApJ...689..919P,2013arXiv1308.0021G}. Additionally, a fast prescription of cluster evolution can be combined with models of galaxy evolution or cosmology.  Both applications are currently out of reach because existing, more sophisticated, methods to solve the $N$-body problem are computationally too expensive  \citep[for a review see the supplementary material of ][]{2010ARA&A..48..431P}. 

\citet{2011MNRAS.413.2509G}  present a simple analytical theory for  the evolution of $N$ and half-mass radius of tidally limited clusters. The model assumes that there is always a balance between the rate of energy generation in the core and the flux of energy through the half-mass radius by two-body relaxation. 
 The theory connects two existing models of Michel H\'{e}non: the  isolated cluster \citep{1965AnAp...28...62H} and the tidally limited cluster \citep{1961AnAp...24..369H}. To connect these models it was assumed  that the energy conduction rate is the same in both models \citep[for a derivation and comparison of these quantities see][]{2011MNRAS.413.2509G}. 
    Numerical $N$-body simulations recently  confirmed the validity of this assumption \citep[][hereafter Paper~I]{2012MNRAS.422.3415A}. 

In Paper~I we present  the first version of a versatile cluster evolution package in the form of the publicly available code Evolve Me A Cluster of StarS (\emacss)\footnote{The code is available from http://github.com/emacss}. It allows a user to define the cluster and tidal field parameters and the code provides the evolution of cluster parameters based on the assumption of balanced evolution.
The evolution  of the number of stars $N$ and half-mass radius $\rh$ of a cluster are obtained by solving two coupled first-order ordinary differential equations, namely $\dot{N}(N,\rh,\Omega)$ and $\rhdot(N,\rh,\Omega)$ with a fourth-order Runge-Kutta integrator. Here $\Omega$ is the angular frequency of the cluster about the centre of the galaxy. Several assumptions had to be made to reduce the evolution of clusters to such a simple model:  relaxation driven escape of stars is the only mechanism that  reduces $N$; the  cluster evolves in a self-similar fashion, such that $\rh$ is a constant times the virial radius $\rv$ (in this case $\rh=\rv$); cluster orbits are circular and the balanced evolution starts after a fixed number of initial half-mass relaxation time-scales $\trh$ and the cluster is not evolved in that first phase.

This paper extends \emacss\ to include the following physical processes: the evolution of the core radius $\rc$ and core density $\rhoc$, the evolution of $N$ and the radii in the unbalanced evolution phase prior to core collapse and the evolution of the ratio $\rh/\rv$. The last ratio depends on the density profile and therefore the concentration of the cluster.  With these new additions, \emacss\ can also evolve clusters that are initially filling the Roche volume and lose a large fraction of their stars prior to core collapse.  In the current version we  assume that all stars have the same mass.

The structure of the paper is as follows: in Section~\ref{sec:framework} we introduce the theoretical framework of the new version of \emacss. In Section~\ref{sec:nbody} we present a suite of direct $N$-body simulations that is compared to \emacss\ and used to implement the new features. In Section~\ref{sec:results} we demonstrate the performance of \emacss\ by comparing it to all $N$-body models and in Section~\ref{sec:conclusion} we present our conclusions and discuss the future steps for \emacss\ that will include a stellar mass function and the  mass-loss of stars.

%%%%%%%%%%%%%%%%%%%%%%%%%%%%%%%%%%%%%%%%%%%%
\section{Framework}
\label{sec:framework}
In this section, we set out the theoretical framework that is used to describe the  evolution of the core radius $\rc$ and core density $\rhoc$ in the unbalanced phase (Section~\ref{ssec:rc}),  the evolution of the other parameters in the unbalanced phase (Section~\ref{ssec:ub}) and the transition to the balanced phase and the evolution of the core   (Section~\ref{ssec:pcc}). The evolution of the half-mass radius in  balanced evolution and the escape rates in both the balanced phase and the unbalanced phase are discussed in Sections~\ref{subsec:mu} and \ref{ssec:escape}, respectively.
We start by introducing  in Section~\ref{ssec:def} the variables, time-scales and definitions used in this paper.

%___________________________________________________________________
\subsection{Variables, definitions and time-scales}
\label{ssec:def}
A fundamental aspect of the evolution of a collisional system, i.e. a star cluster, is  the increase of the total energy  (the system becoming less bound) on a time-scale shorter than the age of the Universe, because of two-body relaxation. For clusters in weak tidal fields, this energy increase (i.e. less negative) results in an expansion of the cluster and for  tidally limited clusters the energy increase results in the escape of stars. The  quantity we want to evolve in a cluster model is, therefore, the total energy $E$ of the cluster (\citealt{2011MNRAS.413.2509G}; Paper~I).
For a  self-gravitating  system in virial equilibrium $E$ can be written as
\begin{equation}
E=-\kappa \frac{GM^2}{\rh}.
\label{eq:e}
\end{equation}
Here, $G$ is the gravitational constant, $M$ and $\rh$ are the mass and the half-mass radius of the cluster, respectively, and $\kappa$ is a form-factor that depends on the density profile of the cluster. In the definition of $E$, we do not include the binding energy of multiple stars. This definition of $E$ is often referred to as the {\it external} energy \citep[as in][]{1997MNRAS.286..709G}.
We assume that the only contributions to the total energy are the kinetic energy $T$ and the gravitational energy $W$, such that $E=T+W = W/2 = -T$. Combined with the definition of the virial radius $\rv=-GM^2/(2W)$ we then find that $\kappa = \rh/(4\rv)$. Note that we ignore the contribution of the tidal field $\etide$ to the total energy. \citet{1995MNRAS.276..206F} show that the ratio $\etide/W\simeq 0.4(\rh/\rj)^3$ for a tidal field due to a point-mass galaxy, which even for very large ratios of $\rh/\rj$ results in a relative contribution of $\etide$ to $E$ of only a few percent. We do include the effect the tides have on the escape of stars.

Taking the time-derivative on each side of equation~(\ref{eq:e}) and dividing by $-E$ we find how the fractional change in energy relates to the fractional change in the other variables
\begin{equation}
-\frac{\edot}{E}= -\frac{\dot{\kappa}}{\kappa} +\frac{\rhdot}{\rh}- 2\frac{\dot{m}}{m}- 2\frac{\dot{N}}{N} .
\label{eq:edot}
\end{equation}
Here we have used $M=mN$, where $m$ is the mean mass of the stars and  $N$ is the number of stars.
In this work we assume single-mass clusters so $\dot{m}=0$ from now on\footnote{The variation of the mean stellar mass as the result of mass-loss from stars and the preferential ejection of low-mass stars will be included in version~3 (Alexander~et~al.~in~preperation).}.
We are interested in the evolution of these quantities on 
a half-mass relaxation time-scale $\trh$ which is defined as  \citep{1971ApJ...164..399S}

\begin{equation}
\trh=0.138\frac{N^{1/2}\rh^{3/2}}{\sqrt{Gm}\ln(0.11N)}.
 \label{eq:trh}
\end{equation}
Here $\ln(0.11N)$ is the Coulomb logarithm and the argument is appropriate for single-mass clusters \citep{1994MNRAS.268..257G}.
To describe the fractional change of the cluster properties per $\trh$ we  define the following dimensionless parameters:

\begin{align}
\epsilon &\equiv -\frac{\dot{E}\trh}{E},\label{eq:epsilon} \\
\lambda &\equiv \frac{\dot{\kappa}\trh}{\kappa},\label{eq:lambda}\\
\mu &\equiv \frac{\rhdot\trh}{\rh}\label{eq:mu},\\
\xie &\equiv - \frac{\dot{N}\trh}{N}\label{eq:xie}.
\end{align}
In  \citet{2011MNRAS.413.2509G} it was assumed that  the dimensionless rate of evolution of energy is constant during the entire evolution, i.e. 
\begin{equation}
\epsilon= \zeta  \simeq 0.1.
\label{eq:zeta} 
\end{equation}
 Here $\zeta$ can be interpreted as the efficiency of energy conduction of the cluster and depends on the stellar mass spectrum in the sense that clusters with a wider mass spectrum evolve faster \citep*{1971ApJ...164..399S,1998ApJ...495..786K}.  
In Paper~I, we used $\epsilon=0$  in the unbalanced phase (energy is conserved), which is accurate for isolated clusters and approximately correct for clusters in weak tidal fields. In this work, we allow for unbalanced evolution of the cluster such that $\epsilon\ge0$ and $\epsilon\ne\zeta$  in the unbalanced phase (Section~\ref{ssec:ub}) and $\epsilon = \zeta$ in the balanced phase (Section~\ref{subsec:mu}). 
  In the unbalanced  phase, $\lambda$ is positive because the cluster gets more concentrated and it is negative in the later evolution. In Paper~I we considered clusters that start deeply embedded within $\rj$ ($\rhrj\equiv\rh/\rj\lesssim1/30$), which means that   $\mu$ is always positive in the initial phase of balanced evolution because the cluster expands to the tidal radius. In (roughly) the second half of the evolution $\mu$ is negative and equals approximately  $-\xie/3$ because the cluster contracts at  a (roughly) constant density in the tidal field   \citep{1961AnAp...24..369H, 2011MNRAS.413.2509G}. 
  In this paper we consider clusters that initially fill the Roche volume  ($\rhrj\simeq0.1-0.2$) and for these clusters $\mu$ can be negative at the start of the evolution.  The value of $\xie$ is always positive, because  $\dot{N}$ is always negative.

If we multiply both sides of equation~(\ref{eq:edot}) by $\trh$ we can   write the evolution of the energy in terms of the dimensionless quantities defined in equations~(\ref{eq:epsilon})-(\ref{eq:xie}), i.e.
\begin{equation}
\epsilon= -\lambda +\mu + 2\xie.
\label{eq:edot2}
\end{equation}
 The reader may have noted that we have not mentioned the core radius $\rc$ so far, whilst we set out to include the evolution of $\rc$ in the model. We have   thus far omitted $\rc$  from the equations because $\rc$ only enters indirectly in the definition of $E$ through $\kappa$, which can be interpreted as a  concentration parameter.  The concentration of a cluster in the well-known \citet{1966AJ.....71...64K} models is defined as the logarithm of the ratio $\rt/\rc$, where $\rt$ is the King truncation radius which is the radius at which the density drops to zero. Here, we make the assumption that throughout the entire evolution $\kappa$ depends only on the ratio $\rcrh\equiv\rc/\rh$, i.e. $\kappa = \kappa(\rcrh)$, independent of the tidal truncation radius. This is motivated by the fact that the total energy is most sensitive to variations of the mass distribution within $\rh$, where the gravitational energy is highest.  
  In Section~\ref{sec:nbody}, we show that results of $N$-body models support this assumption. To proceed, we introduce an  additional dimensionless  parameter 
 \begin{align}
\delta &\equiv \frac{\rcdot\trh}{\rc}
\label{eq:delta}
\end{align}
for the evolution of the core radius $\rc$ on a $\trh$ time-scale.
To include $\delta$ in the energy equation~(\ref{eq:edot2}) we  take the time derivative of $\kappa(\rcrh)$, using $\rcrhdot/\rcrh = \rcdot/\rc - \rhdot/\rh$,
such that

\begin{equation}
\frac{\dot{\kappa}}{\kappa} = \K\left(\frac{\rcdot}{\rc} - \frac{\rhdot}{\rh}\right),
\label{eq:kappadot}
\end{equation}
with $\K \equiv \dr \ln\kappa/\dr\ln\rcrh$.
With this expression we can relate the dimensionless parameter $\lambda$ that describes the evolution of $\kappa$  (equation~\ref{eq:lambda}) to the dimensionless parameters for the half-mass radius and core radius, $\mu$ (equation~\ref{eq:mu}) and $\delta$ (equation~\ref{eq:delta}), respectively,
\begin{equation}
\lambda = \K(\delta - \mu).
\label{eq:lambda2}
\end{equation}
 We substitute this in equation~(\ref{eq:edot2})  to find 
\begin{equation}
\epsilon  = -\K\delta  + (1+\K)\mu +2\xie.
\label{eq:meq}
\end{equation}
This equation relates the evolution of the total energy $E$ to the evolution of the core radius $\rc$ (through $\delta$),  the half-mass radius $\rh$ (through $\mu$) and the  number of stars $N$ (through $\xie$). 
It is this equation we are going to solve to get the time evolution of $\rc$, $\rh$ and $N$ in  the unbalanced phase. 
Before we discuss the change of energy $\epsilon$ in the unbalanced phase in Section~\ref{ssec:ub}, we  first  discuss  the rate at which the core radius contracts  in the unbalanced phase.\\

%___________________________________________________________________
\subsection{Core contraction and gravothermal catastrophe }
\label{ssec:rc}

In the earliest phase of unbalanced evolution of a single-mass cluster the contracting core converts gravitational energy in kinetic energy which provides the energy that is required by two-body relaxation. Because the energy requirement is set by the cluster as a whole the core  contracts on a half-mass relaxation time scale. Because of our definition of $\delta$  (equation~\ref{eq:delta}) it follows that $\delta$ is approximately constant in that phase. 
When the relaxation time-scale of the core itself becomes  much shorter than $\trh$ then a runaway contraction follows. This process is often referred to as core collapse, or the gravothermal catastrophe \citep{1968MNRAS.138..495L} and it takes over from the slow contraction when the core radius  becomes smaller than $\rc\lesssim0.07\rh$   \citep{1980ApJ...242..765C}. From that moment  the evolution of the core is decoupled from the evolution of the cluster and the core contracts self-similarly on a core relaxation time-scale $\trc$  \citep{1980MNRAS.191..483L} until the collapse is halted by the formation of the first hard binary \citep[in the absence of other energy sources, such as primordial binaries, a central black hole or stellar mass-loss, ][]{1975MNRAS.173..729H}. The definition of $\trc$  is \citep{1971ApJ...164..399S}

\begin{equation}
\trc = \frac{\sigmac^3}{15.4G^2m\rhoc\ln(0.11N)}.
\label{eq:trc}
\end{equation}
Here, $\sigmac^2$ is the  mean-square velocity of stars in the core and $\rhoc$ is the core density. 
The core is to good approximation an isothermal system and $\sigmac^2$ can be written as $\sigmac^2=(4/3)\pi G\rhon\rc^2$, where $\rhon\simeq2\rhoc$ is the central density.
During the gravothermal catastrophe the core density increases as 
\begin{equation}
\rhoc =\rhocn \rc^{-\alpha},
\label{eq:rhoc}
\end{equation}
 where $\rhocn$ is a constant of proportionality and $2.2\lesssim\alpha\lesssim2.3$ \citep{1980MNRAS.191..483L,1988MNRAS.230..223H, 2003MNRAS.341..247B}. For simplicity we assume that this relation holds during the entire unbalanced phase so we can write 

\begin{equation}
\sigmac^2 = \frac{8}{3}\pi G\rhocn \rc^{2-\alpha}.
\label{eq:sigmac}
\end{equation}
Now $\trc$ is only a function of one variable ($\rc$) and  two parameters ($\rhocn$ and $\alpha$), which  are determined in Section~\ref{sec:nbody}.
For the rate of core contraction during the gravothermal catastrophe we use   $\delta_2 = \rcdot\trc/\rc$. To ensure a smooth transition between the two different phases we define $\delta$ as

\begin{equation}
\delta = \delta_1 + \delta_2\frac{\trh}{\trc}.
\label{eq:delta2}
\end{equation}
Here $\delta_1$ is a negative constant that describes the speed of the initial contraction on a $\trh$ time-scale and $\delta_2$ is a negative constant that describes the gravothermal catastrophe on a $\trc$ time-scale. 
For clusters that start with $\rc/\rh\gtrsim0.07$ the second term on the right-hand side of equation~(\ref{eq:delta2}) is initially small because $\trh/\trc<<\delta_1/\delta_2$ and therefore $\delta\simeq\delta_1$. Whilst the core contracts at this rate, the ratio $\trh/\trc$ grows and at some point the second term becomes dominant and during the runaway collapse we have $\delta \simeq \delta_2\trh/\trc$. 
Combined with  equation~(\ref{eq:delta}) we find that  in this phase $\rcdot/\rc = \delta_2/\trc$.
In Section~\ref{sec:nbody} we will demonstrate that this simple linear addition of the two core contractions rates accurately describes the evolution of $\rc$ and we determine the constants $\delta_1$ and $\delta_2$ from theory and $N$-body models.

Now that we have defined how $\delta$ depends on the other cluster parameters, we turn to the variation of $\epsilon$ in the unbalanced phase.

%___________________________________________________________________
\subsection{Unbalanced/pre-collapse evolution}
\label{ssec:ub}
To be able to numerically solve equation~(\ref{eq:meq}) we need to have an  expression for the rate of change  of energy $\epsilon$ in the unbalanced phase.  In this phase the cluster has no energy source and the core contracts to generate heat. In isolation, the total energy of the cluster is conserved ($\epsilon=0$, Paper~I). In a tidal field, the energy of the cluster can change because of the escape of stars over the tidal boundary. This is an important effect to consider for clusters in a strong tidal field, because for these clusters more than half of the stars can escape before core collapse \citep[e.g.][]{2001MNRAS.325.1323B}.

For most of the unbalanced phase the escape of stars happens on a relaxation time-scale because the outer parts of the cluster expand while the core contracts  on a $\trh$ time-scale (Section~\ref{ssec:rc}) and the response of the cluster can be implemented with straight forward energy considerations. 
Assume a cluster that has a large ratio $\rhrj\simeq0.1-0.2$, meaning that the cluster `fills' the Roche volume.
 Then assume that  stars  gain energy by relaxation effects until they reach the escape energy and leave the cluster  through the Lagrangian points with small velocities, such that the specific energy of the escaping stars is approximately $-GM/\rj$. 
The change in energy as a result of the loss of stars is thus $\dr E = -(GM/\rj)\dr M$. Dividing this by $E/\trh$ we find that the energy increase depends on the escape rate as 

\begin{equation}
\epsilon = \frac{\rhrj}{\kappa}\xie.
\label{eq:em}
\end{equation}
To understand the cluster's response to the loss of stars, we substitute this expression for $\epsilon$ in equation~(\ref{eq:meq}) and find for the evolution of $\rh$
\begin{equation}
\displaystyle\mu  =  \frac{\left(\rhrj/\kappa-2\right)\xie + \K\delta }{1+\K}.
\label{eq:mu2}
\end{equation}
Because $\trh$ and all the terms on the right-hand side of  equation~(\ref{eq:mu2}) are functions of $\rc$, $\rh$, $N$ and the angular frequency of the cluster about the Galaxy centre $\Omega$, we can rewrite equation~(\ref{eq:mu2}) as $\rhdot(\rc, \rh, N, \Omega)$. This we can solve simultaneously with $\rcdot(\rc, \rh, N)$ and  $\dot{N}(\rc, \rh, N, \Omega)$ with a simple fourth order Runge-Kutta integrator, as in Paper~I.  
To be able to solve these equations in time we need to have an expression for $\xie$, which is the topic of Section~\ref{ssec:escape}.

From equation~(\ref{eq:mu2}) we see that the rate at which a cluster shrinks, or expands, depends critically on the  ratio $\rhrj$. Consider a Plummer model with $\rhrj=5\kappa/3$. Because for this model  $\kappa \simeq 0.2$ we have $\rhrj\simeq 0.333$ and we find that $\mu\simeq-(1/3)\xie$ (ignoring the small contribution of $\K$). This means that the half-mass radius shrinks as $N^{1/3}$ as the cluster loses stars. Because $\rj$ also shrinks as $N^{1/3}$ in response to the escape of stars we find that for this $\rhrj$ the cluster shrinks at a constant density and, therefore, constant $\rhrj$. For $\rhrj\gtrsim1/3$, and under the assumption that the density profile (i.e. $\kappa$) does not change, the cluster is unstable and will go into a runaway dissolution. For  clusters with $\rhrj<5\kappa/3\simeq0.333$  $\rh$ shrinks faster than $\rj$ until an energy source becomes active. 

Clusters in the post-collapse phase evolve roughly  at a constant $\rhrj\simeq0.145$ \citep[][]{1961AnAp...24..369H}, i.e. much lower than $1/3$. This is because the energy of these clusters changes not only because of a loss of stars over the tidal boundary, but also because of energy production in the core \citep[see the discussion on p.~57 of Chapter 3.2 in][]{1987degc.book.....S}.
In the next Section we discuss the transition to the balanced phase.

%___________________________________________________________________
\subsection{Core collapse criterion and core evolution in the balanced phase}
\label{ssec:pcc}

Before we can define the exact condition for the transition from unbalanced to balanced evolution it is necessary that we  consider first the evolution of $\rc$ in the balanced phase.

%______________________________

\subsubsection{Core evolution in the balanced phase}
\label{ssec:coreevol}

In the balanced phase the size of $\rc$ depends on the amount of energy that is produced, which in turn is set by the energy demand of the cluster as a whole (H\'{e}non's principle). For realistic clusters it can get complicated to understand this when we consider the combined effect of (primordial) binary stars, black holes, stellar mass-loss, etc. For single-mass clusters without primordial binary stars, however, it is possible to express the evolution of $\rc$ in terms of $\rh$ and $N$. With the assumption of energy balance and steady heating by binary stars that form in multiple encounters  one can derive that in this phase the core radius depends on $N$ and $\rh$ as \citep[see box 28.1 in][]{2003gmbp.book.....H} $\rc = \left(N/N_2\right)^{-2/3}\rh$,
where $N_2$ is a constant that will be determined in Section~\ref{sec:nbody}. The evolution of $\rc$ is passive, in the sense that it follows the evolution of  $N$ and $\rh$ which follow from the assumption of balanced evolution (\citealt{2011MNRAS.413.2509G}; Paper~I). 

For clusters with $N\gtrsim7000$ there is no steady core evolution, but the core undergoes gravothermal oscillations \citep{1984MNRAS.208..493B, 1987ApJ...313..576G}. We do not include these oscillations of the core, although a simple prescription exists \citep{1992MNRAS.257..245A}. Instead, we assume that for large $N$  the ratio $\rcrh$ tends to a constant $\rcrh\simeq(N_3/N_2)^{-2/3}$, where $N_3\simeq 7000$ is the boundary between clusters for which $\rcrh$ evolves as $N^{-2/3}$ (i.e. for $N\lesssim N_3$) and those for which $\rcrh$ is constant (i.e. for $N\gtrsim N_3$). The exact value for $N_3$ will be determined in Section~\ref{sec:nbody}. To implement the convergence to a constant $\rcrh$ for clusters with large $N$ we use 

\begin{align}
\rcrh &= \left(\frac{N_2}{N} + \frac{N_2}{N_3}\right)^{2/3},\label{eq:rcpost}\\
&\simeq
\begin{cases}
\left(N/N_2\right)^{-2/3}&\text{for $N<<N_3$};\\
\left({N_3}/{N_2}\right)^{-2/3} &\text{for $N>>N_3$}.\\
\end{cases}
\end{align}
Taking the time derivative of equation~(\ref{eq:rcpost}) and multiplying by $\trh$ we find an expression for $\delta$ in the post-collapse phase
\begin{equation}
\delta = \frac{2}{3}\xie\left(1+\frac{N}{N_3}\right)^{-1}+ \mu.
\label{eq:delta3}
\end{equation}
For large $N\gg N_3$ the core radius evolves at the same rate as the half-mass radius because the first term on the right-hand side is negligible and therefore $\delta\simeq\mu$, while for $N\lesssim N_3$ the ratio $\rcrh$ grows as $N^{-2/3}$ while $N$ decreases. The evolution of $\rh$ (i.e. $\mu$) is discussed in Section~\ref{subsec:mu}.

For the evolution of the core density $\rhoc$ we assume that  between $\rc$ and $\rh$ the cluster is approximately  isothermal and has a density distribution $\rho\propto r^{-2}$, such that

\begin{equation}
\rhoc = \rhoh\rcrh^{-2},
\end{equation}
where $\rhoh = 3M/(8\pi\rh^3)$ is the average density within $\rh$.

Now we have defined the equilibrium evolution of $\rc$ and $\rcrh$ in the balanced phase, we consider the transition from unbalanced to balanced evolution.

%______________________________
\subsubsection{Criterion for core collapse}
We  define the moment of core collapse as the moment in the evolution that $\rcrh$   has reached  the value of the relation for $\rcrh$ as a function of $N$ in the balanced phase (equation~\ref{eq:rcpost}). At each time step in the unbalanced phase the criterion  changes because it depends on the instantaneous value of $\rcrh$ and $N$. This allows us to make the transition to the balanced evolution without a priori (i.e. before the evolution starts) knowledge of the exact {\it moment}  of core collapse. Core collapse time is well understood for isolated, single-mass, Plummer models: roughly $17$ initial $\trh$  \citep*[e.g.][]{1970MNRAS.150...93L, 1974A&A....37..183A}, but it is hard to predict what it is when the cluster loses a significant number of stars in the unbalanced phase, or starts with a smaller core. Both effects are now included in the  \emacss\ model. The way we make the transitions causes us to underestimate the maximum core density in the collapse. This is because after core collapse the core expands towards larger radii and this core bounce  \citep{1983MNRAS.205..913I} is not included in the model.
This effect can be seen in the $N$-body models (see Section~\ref{sec:nbody}). The relation we propose describes the evolution of $\rc$ near the maxima after core bounce and is therefore a reasonable description for the majority of the evolution.

%___________________________________________________________________
\subsection{Half-mass radius in balanced evolution}
\label{subsec:mu}
Combining equation~(\ref{eq:delta3}) with the relation for the total energy variation (equation~\ref{eq:meq}) we find that the half-mass radius evolution in balanced evolution relates to $\zeta$ and $\xie$ as
\begin{equation}
\mu =  \zeta + \left(\frac{2}{3}\K\left[1+\frac{N}{N_3}\right]^{-1}-2\right)\xie.
\label{eq:muub}
\end{equation}
 If we ignore the variation of the density profile due to the evolution of $\rc$ (i.e. $\K=0$)  we find $\mu = \zeta -2\xie$, i.e. the relation that was used in Paper~I. 
The small $\K$ dependent term in equation~(\ref{eq:muub}) is the only difference with the radius evolution in the balanced phase presented in Paper~I.
The consequence of this difference is that the evolution of $\rh$ and $\rv$ is slightly different in the balanced phase for clusters with $N\lesssim N_3$, whereas in Paper~I we assumed $\rh/\rv = 1$. In the next section, we discuss the escape rate $\xie$ in both the unbalanced and the balanced evolution.

%___________________________________________________________________
\subsection{Escape rate}
\label{ssec:escape}
Up to this point, we have expressed the evolution in terms of $N$ and the dimensionless escape rate   $\xie$. To be able to solve all relations in time, we need an expression for $\xie$ and the initial number of stars $N$. In this section, we find expressions for $\xie$ in the balanced phase (Section~\ref{sssec:xie_u}) and in the unbalanced phase (Section~\ref{sssec:xie_b}).
From the $N$-body simulations (Section~\ref{sec:nbody}), we find that $\xie$ in the unbalanced phase is lower than what we found for the balanced evolution in Paper~I1. An increase of the mass-loss rate after core collapse was also found for multimass model by \citet*{2010MNRAS.409..305L}. Before we can describe $\xie$ in the unbalanced phase, we need to first recall  the definition of $\xie$ in the balanced phase as described in detail in Paper~I.

%______________________________
\subsubsection{Escape rate in the balanced phase}
\label{sssec:xie_u}
In this section, we discuss the escape rate of stars in the balanced phase by recalling the framework described in Paper~I.
The arguments used in Paper~I follow from the results of \citet{2008MNRAS.389L..28G}, who find that the escape rate in $N$-body models of tidally limited clusters depends on the ratio $\rhrj$ and $N$ as $\xie\propto\rhrj^{3/2} N^{1/4}$. The scaling $\rhrj^{3/2}$ is because the escape energy is lower for larger $\rhrj$, which makes it easier for a larger fraction of the stars to escape in a $\trh$ time-scale. The scaling with $N^{1/4}$ is because of the delayed escape of stars from the anisotropic Jacobi surface \citep{2000MNRAS.318..753F}, which preferentially slows down the escape of stars from low-$N$ systems \citep{2001MNRAS.325.1323B}. Isolated clusters lose a small fraction (approximately a percent) of their stars every relaxation time \citep{2002MNRAS.336.1069B}. To include both effects, we used the following expression for $\xie$ in Paper~I

\begin{equation}
\xie =\xieo(1-\PP) + \frac{3}{5}\zeta \PP,
\label{eq:xieb}
\end{equation}
where $\xieo = 0.0142$ (Paper~I) is the escape rate  for isolated clusters and 

\begin{equation}
\PP = \left( \frac{\rvrj}{\rvrjo} \right)^{z}\left( \frac{N}{N_1}\frac{\log [0.11 N_1]}{\log[ 0.11 N]} \right)^{1-x},
\label{eq:P}
\end{equation}
with $z = 1.61$ (Paper~I), $x = 0.75$ (\citealt{2001MNRAS.325.1323B}; Paper~I) and  $\rvrjo = 0.145$ (\citealt{1961AnAp...24..369H}; Paper~I). In weak tidal fields $\PP\simeq 0 $ and $\xie \simeq \xieo$, a constant rate of escape per relaxation time, while for tidally limited clusters the quantity $\PP \simeq 1 $ and $\xie \simeq (3/5)\zeta\simeq0.06$.  
The scaling constant $N_1$ was determined in Paper~I ($N_1=38\,252$), but in Section~\ref{sec:nbody} we slightly revise this value. This is because in equation~(\ref{eq:P}) we use $\rv$ in the ratio $\rvrj$ and $\trh$ is expressed in terms of $\rh$ and in the current version $\rh/\rv$ is allowed to evolve, whereas in Paper~I $\rh$ was always equal to $\rv$.

%______________________________
\subsubsection{Escape rate in the unbalanced phase}
\label{sssec:xie_b}
The expression for $\xie$ in the unbalanced phase should satisfy three conditions: first, isolated clusters lose almost no stars \citep{2002MNRAS.336.1069B}; secondly, the escape rate of Roche volume filling clusters is about $f\simeq0.3$ times that in the balanced phase and, finally, it should connect to $\xie$ in the balanced phase. 
We therefore adopt the following relation for $\xie$ in the unbalanced phase

\begin{align}
\xie &= \F\xieo(1-\PP) + (f+[1-f]\F)\frac{3}{5}\zeta \PP,\label{eq:xieub}\\
&=
\begin{cases}
\hspace{1.65cm}f(3/5)\zeta\PP&\text{for $\F = 0$};\\
\xieo(1-\PP) + (3/5)\zeta \PP &\text{for $\F = 1$}.\\
\end{cases}
\end{align}
Here $\F = \rcrhmin/\rcrh$ and $\rcrhmin$   is  the minimum ratio of $\rcrh(N)$ in the unbalanced phase and is reached at the moment of  core collapse (equation~\ref{eq:rcpost}).
In the beginning of the evolution of low-concentration clusters (such as Plummer models), we have $\rcrh\gg\rcrhmin$ and therefore $\F\simeq 0$ and  there is only a contribution from escapers due to the tidal truncation: $\xie \simeq f (3/5)\zeta$. This relation ensures that $\xie\simeq 0$ for isolated clusters in the unbalanced phase, as it should. Close to core collapse $\rcrh\simeq\rcrhmin$ and therefore $\F\simeq 1$ such that both the term due to escapers in isolation and the term due to escapers in the tidal field  approach the values in  balanced evolution. 

In the next section we discuss the implementation of these equations in \emacss\ and a comparison to $N$-body simulations.

%%%%%%%%%%%%%%%%%%%%%%%%%%%%%%%%%%%%%%%%%%%%
\section{Implementation and comparison to $N$-body simulations}
\label{sec:nbody}

%___________________________________________________________________
\subsection{Description of $N$-body simulations}
Here we describe the details of a suite of direct $N$-body simulations to benchmark the \emacss\ model against. 
 We model clusters  with five different values of $N$ ranging from  $N= 4\,096$ to  $65\,536$ with steps of a factor of two. All stars have the same mass and the clusters were initially described by \citet{1911MNRAS..71..460P} models or \citet{1966AJ.....71...64K} models with $W_0=5$ with isotropic velocity distributions. 
  The latter model was used for the simulations of clusters in strong tidal fields to avoid having stars above the escape energy.
 We  used the standard $N$-body units, such that $G=M=-4E=1$ \citep{1986LNP...267..233H}. The virial radius $\rv$ is defined as $\rv = -GM^2/(2W)$, where $W$ is the gravitational  energy. We assume  that the clusters are in virial equilibrium initially, such that $W = 2E$ and $\rv = 1$. In this case, the conversion factor for time 
in physical units ($t^*$) relates to the value of $\rv$ in physical units ($\rv^*$) and the mass in physical units ($M^*$) as $t^* = (GM^*/{\rv^*}^3)^{-1/2}$.
 The half-mass radii for the Plummer and King models in these units are $\rh\simeq0.78$ and $0.82$ respectively. The initial value for $\kappa$ for the two models is thus $\kappa_0 \simeq 0.195$ and $0.205$. In \emacss\ $\kappa_0$ is computed from the initial $\rh$ as $\kappa_0 = \rh/4$ (because $\rv=1$).
 
 The equation of motion of the stars was solved in a reference frame that corotates with the circular  orbit of the cluster about  a point-mass  galaxy. The centrifugal, Coriolis and tidal forces were added to the forces due  to the other $N-1$ stars \citep[equation~1 in][]{1997MNRAS.286..709G}. The strength of the tidal field can be quantified by the angular frequency $\Omega$ of the cluster orbit. For a circular orbit around a point-mass galaxy the Jacobi radius $\rj$ of the cluster depends on $\Omega$ and the mass of the cluster $M$ as

\begin{equation}
\rj =\left(\frac{GM}{3\Omega^2}  \right)^{1/3}.
\label{eq:rj}
\end{equation}

We modelled four sets of clusters with  different initial ratios $\rhrj$. Two sets of compact (in terms of $\rhrj$) clusters were presented in Paper~I. These clusters were initially described by Plummer models and the two sets had initial values of $\rhrj=1/100$ and $1/30$.
For this study we ran two additional sets of `Roche filling' clusters with $\rhrj=1/10$ (Plummer) and a series of \citet{1966AJ.....71...64K} models with $W_0 = 5$. For the latter set of runs $\rhrj = 1/5.37$, but we will refer to these runs as $\rhrj=1/5$. For low-$N$ clusters multiple simulations were done to average out statistical fluctuations, in the same way as was done in Paper~I:  for $N = [4\,096, 8\,192,16\,384,32\,768,65\,536]$ we ran $[16,8,4,2,1]$ simulations, respectively.

Stars are  counted as members when their distance to the centre of the cluster is less than $\rj$ and stars are removed from the simulation if their distance from the cluster centre exceeds $2\rj$. The Jacobi radius $\rj$ and the number of members are  calculated iteratively using equation~(\ref{eq:rj}). The core radius is defined as in chapter 15.2 of  \citet[][]{2003gnbs.book.....A} and with this definition for $\rc$ both the  Plummer model and the King model with $W_0 = 5$ have $\rcrh\simeq0.4$.
The energy $E$ of the cluster is defined as the external energy \citep[kinetic and potential components of single stars and the centres of mass of multiples, see][]{1997MNRAS.286..709G} separately from the ÔinternalÕ energy of particles (i.e. the energy  stored in binaries and multiples). For all simulations we used the $N$-body code {\small\textsc{NBODY6}}, which is a fourth order Hermite integrator with \citet{1973JCoPh..12..389A} neighbour scheme \citep{1992PASJ...44..141M, 1999PASP..111.1333A, 2003gnbs.book.....A} with accelerated   force calculation on NVIDIA Graphics Processing Units \citep{2012MNRAS.424..545N}. In the next sections, we compare the results of the $N$-body models to \emacss\ and determine the parameters. To do this, we isolate the various physical process and build up the model piece by piece in Sections~\ref{ssec:kappan} to \ref{ssec:escub} to find the values of the parameters of the various physical processes described in Section~\ref{sec:framework}. The fluctuations that occur in small $N$ systems are taken into account by comparing \emacss\ to the average of the results for the individual runs with the same initial condition, but different random seeds. The final best-fitting parameters of \emacss\ are summarized in Table~\ref{tab:parameters}. 

%___________________________________________________________________
\subsection{Relation between $\kappa$ and $\rcrh$}
\label{ssec:kappan}
\begin{figure}
\includegraphics[width=8cm]{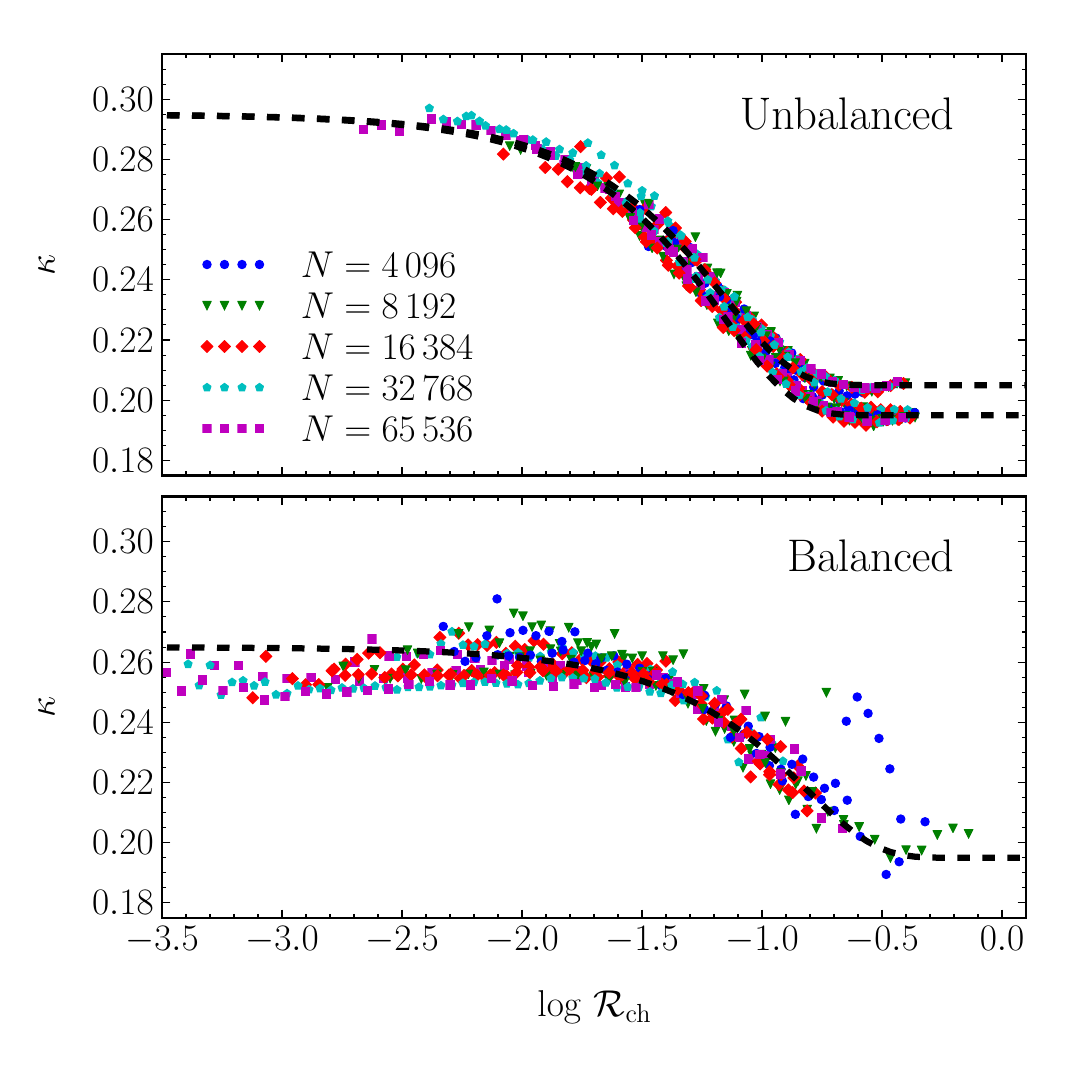}
\caption{The evolution of $\kappa$ as a function of the ratio $\rcrh=\rc/\rh$ for clusters in the unbalanced (pre-collapse) phase (top) and in the balanced (post-collapse) phase (bottom). The dashed lines  approximate the $N$-body results with error functions (equation~\ref{eq:kapparcrh}). In the unbalanced phase there are two dashed lines shown: the top line corresponds to the \citet{1966AJ.....71...64K} models ($\kappa_0\simeq0.205$) and the bottom line corresponds to the \citet{1911MNRAS..71..460P} models ($\kappa_0\simeq0.195$).}
\label{fig:krcrh}
\end{figure}

The first thing we determine from the $N$-body simulations is the relation between $\kappa$ and the ratio $\rcrh$ (Section~\ref{ssec:def}).The points were computed as follows: for 20 runs with $N$ ranging from  $N=4\,096$ to $65\,536$ with steps of 2, and $\rhrj=1/5,1/10,1/30$ and $1/100$ we determined the values of $\kappa$ and $\log\,\rcrh$ from the individual simulations. 
 All runs follow similar tracks, but the relation $\kappa(\rcrh)$  in the unbalanced phase  is different from the relation in the balanced phase. In the balanced phase, there is an indication that $\kappa$ is smaller for 
larger $N$ models  at low values of $\rcrh$, but we will not  include this small $N$ dependence in the model. The difference between the unbalanced and balanced curves is most likely due to the difference in density profile: in the  unbalanced phase the cluster starts with a large core and during the collapse it develops an $r^{-2.2}$  cusp in the central density profile. In the balanced phase, the central density profile is almost isothermal and the central density cusp is  $r^{-2}$. 
Because  of this difference, we describe the $\kappa(\rcrh)$ relation in the different phases with different functions.  
To separate the evolution in the two phases we have to find a definition of core collapse in these models.
We define core collapse as the moment when the total energy $E$ increases by more than 5\% in a unit of $N$-body time. Such a sharp increase in $E$ is not found at any other moment in all runs and turns out to be a useful definition for all simulations. For both the balanced and the unbalanced phase, the median of $\kappa$  was found in 50 bins that were equally spaced in $\log\,\rcrh$. A minimum of $N=200$ remaining stars was used.
 In Fig.~\ref{fig:krcrh}, we show the $\kappa$ values of the $N$-body models as dots with the results for the unbalanced and the balanced phase in the top and bottom panels, respectively.

We find that for both evolutionary phases  the $\kappa$ values can be well described by an error function of the form
\begin{equation}
 \kappa(\rcrh) = \kappa_1+(\kappa_0 - \kappa_1)\,\erf\left( \frac{\rcrh}{\rcrhn}\right).
 \label{eq:kapparcrh}
\end{equation}
The values for the constants are given in Table~\ref{tab:parameters} and we note that for the unbalanced phase $\kappa_0 = \rh/(4\rv)$ depends on the initial density profile of the cluster.
For this function the logarithmic derivative $\K$ (equations~\ref{eq:kappadot} \& \ref{eq:lambda2}) is

\begin{equation}
\K =\frac{\rcrh}{\kappa}\frac{2(\kappa_0-\kappa_1)\exp(-\rcrh^2/\rcrhn^2)}{\sqrt{\pi}\rcrhn}
\end{equation}
and for the parameters used here we find $-0.1\lesssim \K < 0$.

The last point of consideration  is the connection between $\kappa(\rcrh)$ in the unbalanced phase and $\kappa(\rcrh)$  in the balanced phase. Because the constant $\kappa_1$ is different in these two phases the function $\kappa(t)$ is discontinuous at core collapse if we simply jump to the new $\kappa(\rcrh)$ relation at core collapse. This would also result in  a discontinuity  in the energy evolution, which is not desirable. We therefore add a term to $\lambda$ in the balanced phase that ensures that $\kappa(t)$ is continuous and that $\kappa$ evolves to the relation $\kappa(\rcrh)$ of equation~(\ref{eq:kapparcrh}) with the parameters appropriate for the balanced phase on a $\trh$ time-scale. The functional form for $\lambda$ we use in the balanced phase is

\begin{equation}
\lambda = \K(\delta - \mu) + \frac{\kappa(\rcrh) - \kappa}{\kappa(\rcrh)}.
\label{eq:lambda3}
\end{equation}
At the start of balanced evolution (i.e. at core collapse) $\kappa$ is higher than  $\kappa(\rcrh)$, such that the added term on the right-hand side of equation~(\ref{eq:lambda3}) is negative. The difference between $\kappa$ and $\kappa(\rcrh)$ gets smaller every integration step and $\kappa$  approaches  $\kappa(\rcrh)$ asymptotically. We do not include this extra term in the energy balance (equation~\ref{eq:mu2}) such that the system is slightly out of balance,  in the sense that $\epsilon\gtrsim\zeta$,  for a fraction of a relaxation time after core collapse. This phase can be interpreted as the  `core bounce'  phase \citep{1983MNRAS.205..913I} in which excess energy  is released by the newly formed binary star(s) which is diffused by two-body relaxation.

The $\K$ values are quite low and $\K$, therefore, affects the evolution only mildly. In Paper~I, we ignored the variation of $\kappa$ and we assumed that $\rh=\rv$ throughout the evolution. For the $N$-body models, initially $\rh\simeq 0.8\rv$ and the evolution of $\kappa$ in the unbalanced phase causes the ratio $\rh/\rv$ to grow to approximately unity (Section~\ref{ssec:rcnbody}). If $N$ becomes smaller than a few thousand the cluster evolves to low concentration again.

%___________________________________________________________________
\subsection{Evolution of the core parameters}
\label{ssec:rcnbody}
\begin{figure}
\includegraphics[width=8cm]{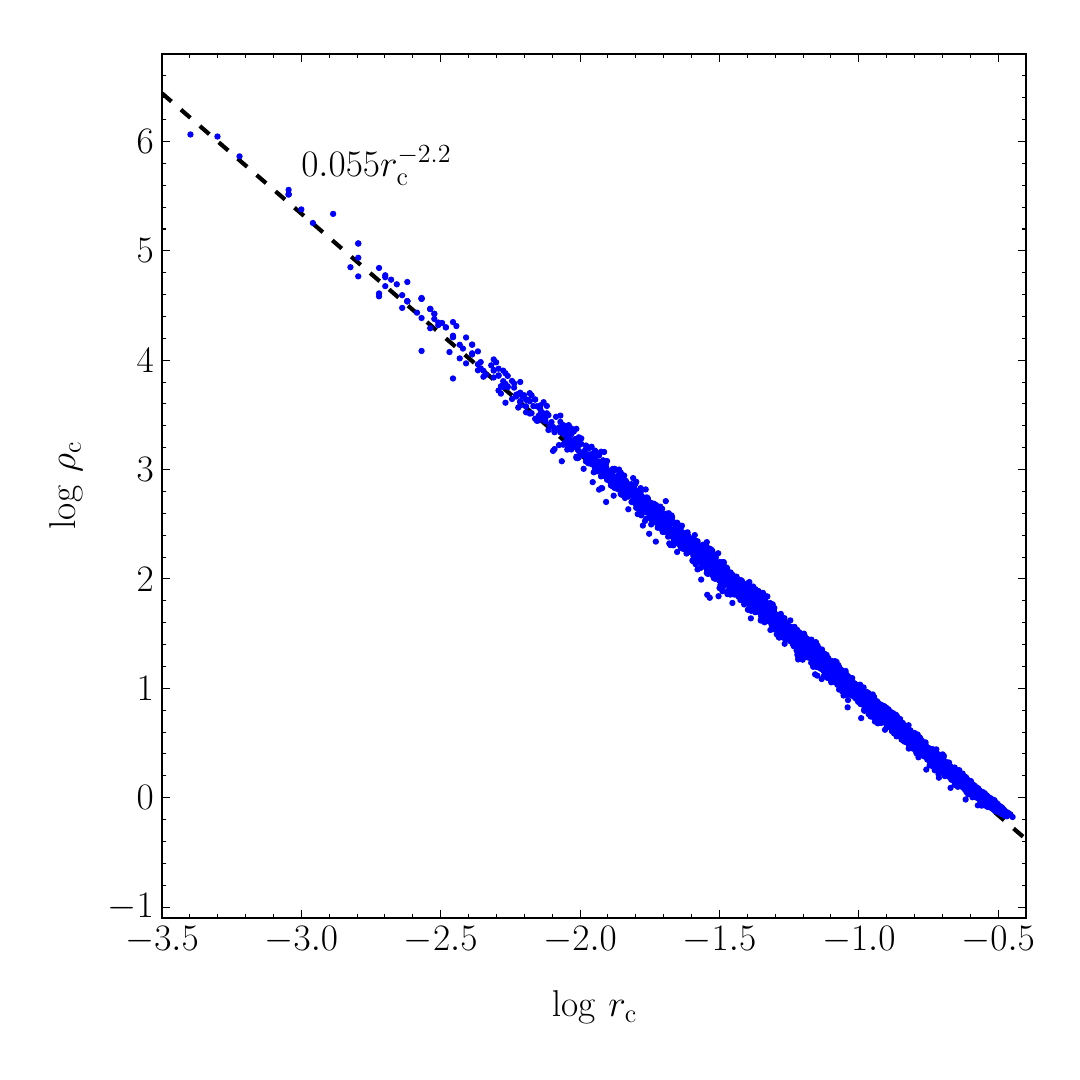}
\caption{Relation between core (mass) density $\rhoc$ and core radius $\rc$ in the unbalanced phase for $N$-body models for the $\rhrj=1/5$ models and the $\rhrj=1/100$ models. For each of the $\rhrj$ sets a model for each $N$ is shown. The tight correlation in the $N$-body data justifies a single relation for $\rhoc(\rc)$ for all models.
The line shows the relation of equation~(\ref{eq:rhocrc}). }
\label{fig:rhocrc}
\end{figure}

To quantify the rate of core contraction we first consider the evolution of the core parameters that define the core relaxation time-scale $\trc$ (equation~\ref{eq:trc}). In Fig.~\ref{fig:rhocrc}, we show the average density within the core $\rhoc$ as a function of $\rc$ in the unbalanced phase for clusters with various initial $N$. The average core density $\rhoc$ is defined as $3\mc/(4\pi\rc^3)$, where $\mc$ is the total mass of the stars in the core.
At the start of the evolution all models start with $\rhoc\simeq 0.7$ and $\rc\simeq 0.3$. When $\rc$ shrinks the density increases   as 

\begin{equation}
\rhoc = 0.055\rc^{-2.2},
\label{eq:rhocrc}
\end{equation}
which corresponds to the dashed line in Fig.~\ref{fig:rhocrc}. 
This value of $\alpha=2.2$ is  close to what was found in previous studies. 
\citet{1980MNRAS.191..483L} used theoretical arguments for the self-similar evolution of the core near the gravothermal catastrophe and found  $\alpha=2.21$.
\citet{1988MNRAS.230..223H} found a logarithmic slope of $-2.23$ from Fokker--Planck models of the late stages of core collapse and   \citet{2003MNRAS.341..247B} found a value of $-2.26$ from $N$-body models of single-mass clusters.
We note that our slightly smaller value of $-2.2$ is probably because we use this relation to describe the entire core contraction phase starting at $t=0$, while the studies mentioned above determined $\alpha$ in the final stages of core contraction (the gravothermal catastrophe).
 With equation~(\ref{eq:rhocrc})  and  the expression for the central velocity dispersion $\sigmac$ (equation~\ref{eq:sigmac}) we have all parameters of the core defined to be able to define $\trc$ (equation~\ref{eq:trc}).

\begin{figure*}
\includegraphics[width=16cm]{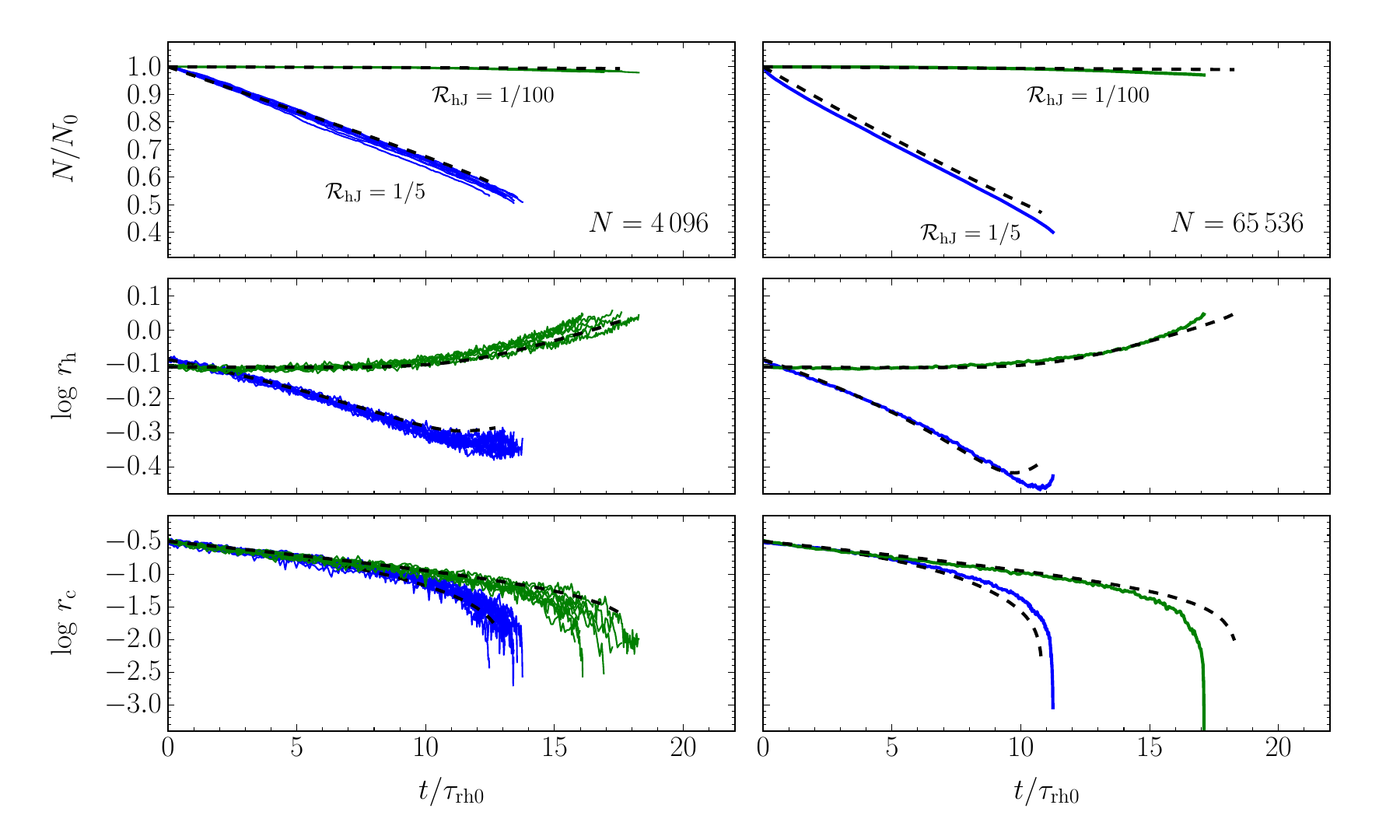}
\caption{Evolution of $N$ (top row), the  half-mass radius $\rh$ (middle row) and the core radius $\rc$ (bottom row) in the unbalanced phase for clusters with different initial $N$:  the 16 different realizations of the $N=4\,096$ model are shown in the left-hand panels and the $N=65\,536$ models  are shown in the right-hand panels. Each panel shows the results for $\rhrj=1/5$ (blue, bottom) and $\rhrj=1/100$ (green, top). The dashed lines show the result of \emacss\ based on the expressions for the evolution of $\rc$ ($\delta$, equation~\ref{eq:delta}) and $\rh$ ($\mu$, equation~\ref{eq:mu2}). }
\label{fig:rc}
\end{figure*}

Now we have defined how $\rhoc$ depends on $\rc$ we can turn to the evolution of $\rc$.
Fig.~\ref{fig:rc} shows the evolution of $N$, $\rh$ and $\rc$  as a function of time following from $N$-body models, expressed in the initial $\trh$, for clusters with different $N$ and $\rhrj$.
In the left-hand panels, the results for clusters with $N=4\,096$ are shown and in the right-hand panels, we show the results for $N=65\,536$. Each panel contains results for clusters with $\rhrj=1/5$ and $1/100$.
The data points were selected to be in the unbalanced phase (pre-collapse) in the same way as described in Section~\ref{ssec:kappan}.

Initially, the core radius shrinks exponentially (i.e. a straight line in logarithmic-linear plot), which is because of the contraction on a $\trh$ time-scale. We find that $\delta_1\simeq-0.09$ (see equation~\ref{eq:delta2}) describes the initial core contraction of the $N$-body models very well. For the clusters with $\rhrj=1/100$ the core radius evolution accelerates after about 15 initial $\trh$ and   $\rc$ contracts on a $\trc$ time-scale. This happens earlier for the $\rhrj=1/5$ clusters because $\trh$ shrinks because of the escaping stars (top panels) and the shrinking $\rh$ (middle panels). 

For the rate of runaway collapse ($\delta_2$), we find that a value of $\delta_2 = 0.002$ provides a good description.
From Fokker-Planck models \citet{1980ApJ...242..765C} finds that in this phase the core density increases at a rate 
$\rhocdot\trc/\rhoc\simeq0.0036$ and \citet{2003MNRAS.341..247B} find $\rhocdot\trc/\rhoc\simeq0.003$ from $N$-body models.
Because of the self-similar nature of the collapse ($\rhoc\propto\rc^{-\alpha}$) we can relate this parameter to $\delta_2$ (equation~\ref{eq:delta2})  as $\delta_2 = -\alpha^{-1}\rhocdot\trc/\rhoc$ (see equation~\ref{eq:delta2}, such that the results of \citet{1980ApJ...242..765C} and \citet{2003MNRAS.341..247B} translate into $\delta_2 \simeq 0.0016$ and $\delta_2\simeq0.0014$, respectively.  It is not a  concern that we need a slightly larger value for $\delta_2$ to get a good description of $\rc$, because \emacss\ does not evolve $\rc$ to the same small values as the Fokker-Planck and $N$-body models, because we switch to balanced evolution once $\rcrh$ reaches the value of balanced evolution (Section~\ref{ssec:pcc}). 

The middle panels of Fig.~\ref{fig:rc} show the evolution of $\rh$. For the clusters with $\rhrj=1/100$ the increase of $\rh$ from  $\rh\simeq0.78$ initially to $\rh\simeq 1$ at core collapse is due to the changing density profile which was already seen in the increase of $\kappa$ (Fig.~\ref{fig:krcrh}). When escaping stars can be ignored the rate of increase of $\rh$ relates to the rate of change of $\rc$ as $\mu=\K\delta/(1+\K)$  (equation~\ref{eq:meq}). This relation nicely describes both the evolution of $\rc$ and $\rh$, also in the presence of escapers as can be seen for the models with $\rhrj=1/5$. We find that \emacss\ slightly underestimates the moment of core collapse for the $\rhrj=1/5$ models and  \emacss\ overestimates this moment for the $\rhrj=1/100$ models. The differences in core collapse times  are in all cases less than approximately $6\%$.

The top panels of Fig.~\ref{fig:rc} show that the clusters in strong tidal fields ($\rhrj = 1/5$)  lose more than half their stars before core collapse. The escape rate $\xie$ in this phase is discussed in more detail in the next section.

%___________________________________________________________________
\subsection{Escape rate}
\label{ssec:escub}

\begin{figure}
\includegraphics[width=8cm]{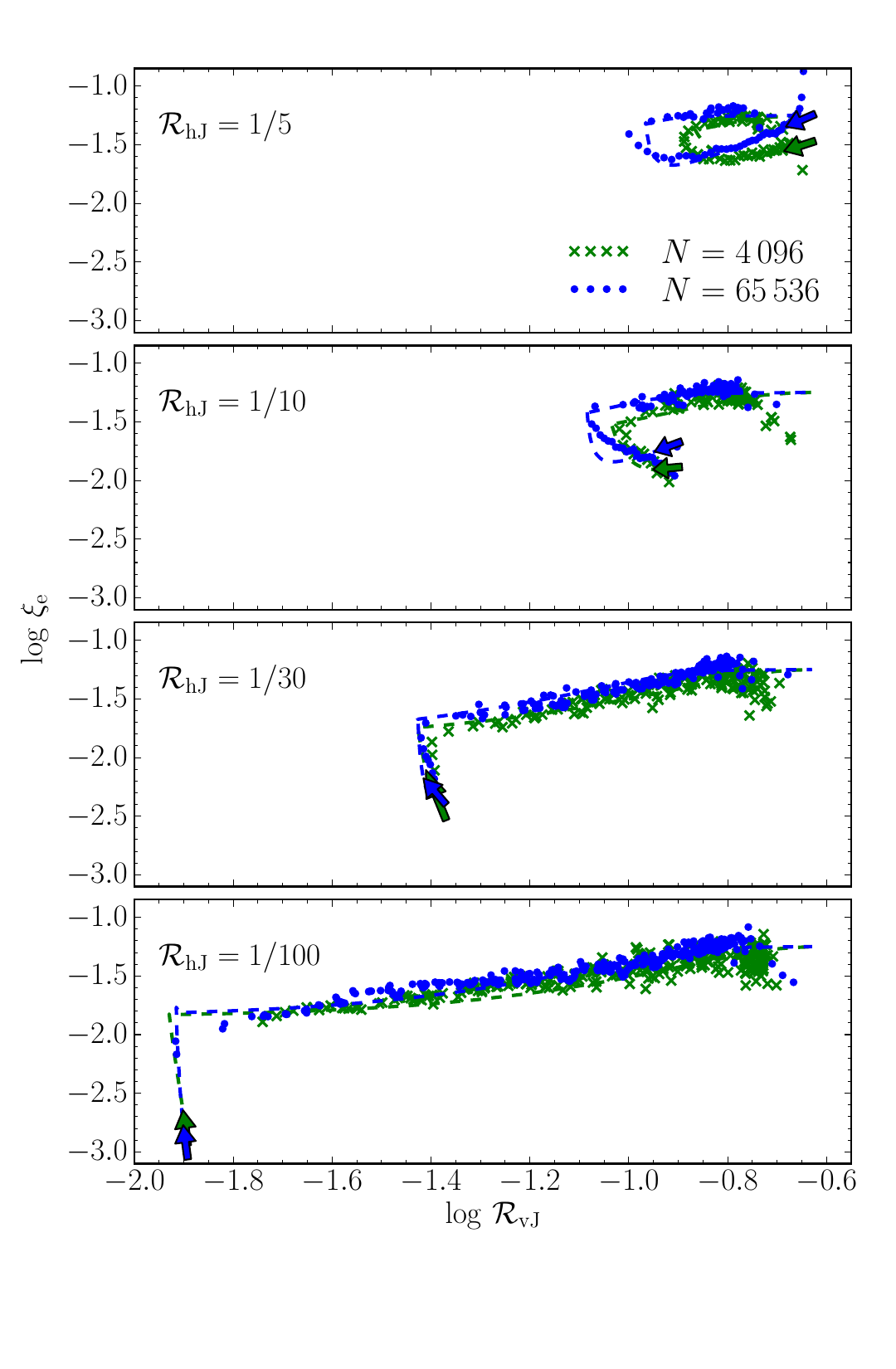}
\caption{Dimensionless escape rate $\xie$ for clusters with $N=4\,096$ (green crosses) and $N=65\,536$ (blue dots) for different initial $\rhrj$.
The derivative $\dot{N}$ for the $N$-body data was found numerically  from the average $N(t)$ data  by dividing the evolution of $N$ in approximately 100 equal steps $\Delta N$ which were  divided by  the corresponding steps $\Delta t$. The start and the direction of the evolution of the \emacss\ result is indicated with an arrow. The dashed lines are the results from \emacss\ using equation~(\ref{eq:xieub}) for the unbalanced phase which connects to the relation for the balanced phase of Paper~I (equation~\ref{eq:xieb}). 
Clusters in a strong tidal field (two top panels) initially contract, i.e. the ratio $\rvrj$ becomes smaller and the cluster moves to the left in this figure. 
Clusters in weak tidal fields (bottom two panels) lose very few stars in the unbalanced phase. Towards  core collapse $\xie$ increases until it reaches the balanced track (Paper~I and equation~\ref{eq:xieb}) where the $\xie$ curves show a sharp bend of approximately 90$^\circ$. }
\label{fig:xie}
\end{figure}

In Fig.~\ref{fig:xie}, we show the dimensionless escape rate $\xie$ as  a function of $\rvrj$ for the entire evolution of clusters with different $N$ ($N = 4\,096$ and $65\,536$) for the four different initial $\rhrj$. The dependence of $\xie$ on $N$ and $\rvrj$ is well described by equation~(\ref{eq:xieub}) and an escape rate due to the tides in the unbalanced phase that is three times lower than what it is in  the balanced phase (i.e. $f=0.33$). When the clusters reach the balanced phase the $\xie$ curves turn by about 90$^\circ$ and the subsequent evolution and corresponding $N$ and $\rvrj$ dependence is well described by the relation from Paper~I  (equations~\ref{eq:xieb} and \ref{eq:P}).

%___________________________________________________________________

\subsection{Integration steps}
In Paper~I, we adopted an integration step of $\Delta t = 0.1\trh$. Here, we need to take smaller steps in the unbalanced phase when the core shrink on a $\trc$ time-scale. We therefore use in the unbalanced phase 

\begin{equation}
\Delta t = \left[(100\trc)^{-1} + (0.1\trh)^{-1}  \right]^{-1}.
\end{equation}
For  small $\trc$ near core collapse  the step size is $\Delta t\simeq 100\trc$. A step size of $100\trc$ is justified by the fact that the  core parameters vary only by a fraction of a per cent near core collapse (Section \ref{ssec:rcnbody}) and with this step size we therefore still under-sample the evolution of the core. 
 A convergence test showed that the final results change by less than 1\% if we decrease  $\Delta t$ by a factor of 100.
In the balanced phase we use $\Delta t = 0.1\trh$, as in Paper~I. \emacss\ outputs the data every $0.1\trh$.

\begin{table}
\caption{Summary of all the parameters  in \emacss}
\label{tab:parameters}
\begin{center}
\begin{tabular}{lcccc}
\hline
Process                               & Quantity              &Unbalanced                  &Balanced           &Equation  \\\hline
Energy diffusion                 &$\zeta$                  & 0.1                              &0.1 $^\dagger$   & (\ref{eq:zeta}) \\
$\rc$ evolution                   &  $\delta_1$            & $-0.09^\ddagger$       &                          & (\ref{eq:delta},\ref{eq:delta2})      \\
                                          & $\delta_2$            & $-0.002^\ddagger$      &                         & (\ref{eq:delta},\ref{eq:delta2})  \\
                                          &$\rhocn$               & 0.055                           &                          & (\ref{eq:rhoc})\\
                                          &$\alpha$                & 2.2                               &                          & (\ref{eq:rhoc})\\
 Concentration                   &$\rcrhn$                & 0.100                           & 0.220                & (\ref{eq:kapparcrh})\\
                                          &$\kappa_0$          & $\rh/(4\rv)$                  & 0.200                  & (\ref{eq:kapparcrh})\\
                                          &$\kappa_1$         & 0.295                           & 0.265                  & (\ref{eq:kapparcrh})\\
 Escape rate $\xie$                   &$f$                       & 0.3                              &                             &(\ref{eq:xieub}) \\  
                                           &$\xieo$               & 0.0142                          &0.0142$^\star$    &(\ref{eq:xieb},\ref{eq:xieub})\\
                                           &$x$                     & 0.75                              &0.75$^\star$        &(\ref{eq:P})\\
                                          &$z$                     & 1.61                               &1.61$^\star$        &(\ref{eq:P})\\
                                          &$\rvrjo$                 & 0.145                            &0.145$^\star$     & (\ref{eq:P})\\
                                           &$N_1$                 &15000                           &15000$^\dagger$ & (\ref{eq:P})\\
$\rcrh$ evolution                &$N_2$                  &                                     &12                       & (\ref{eq:rcpost}) \\
                                          &$N_3$                   &                                    &$15000^\dagger$ & (\ref{eq:rcpost})\\
\hline
\end{tabular}
\end{center}
{ {\it Notes}: $\dagger$ the values found in Paper~I are slightly adjusted; $\ddagger$ in the code these values are normalised to $\zeta$, such that the user can choose to use a different value of $\zeta$ and adjust the speed of the entire evolution; $\star$ from Paper~I.}
\end{table}

%%%%%%%%%%%%%%%%%%%%%%%%%%%%%%%%%%%%%%%%%%%%
\section{General results}
\label{sec:results}
\begin{figure*}
\center\includegraphics[width=16cm]{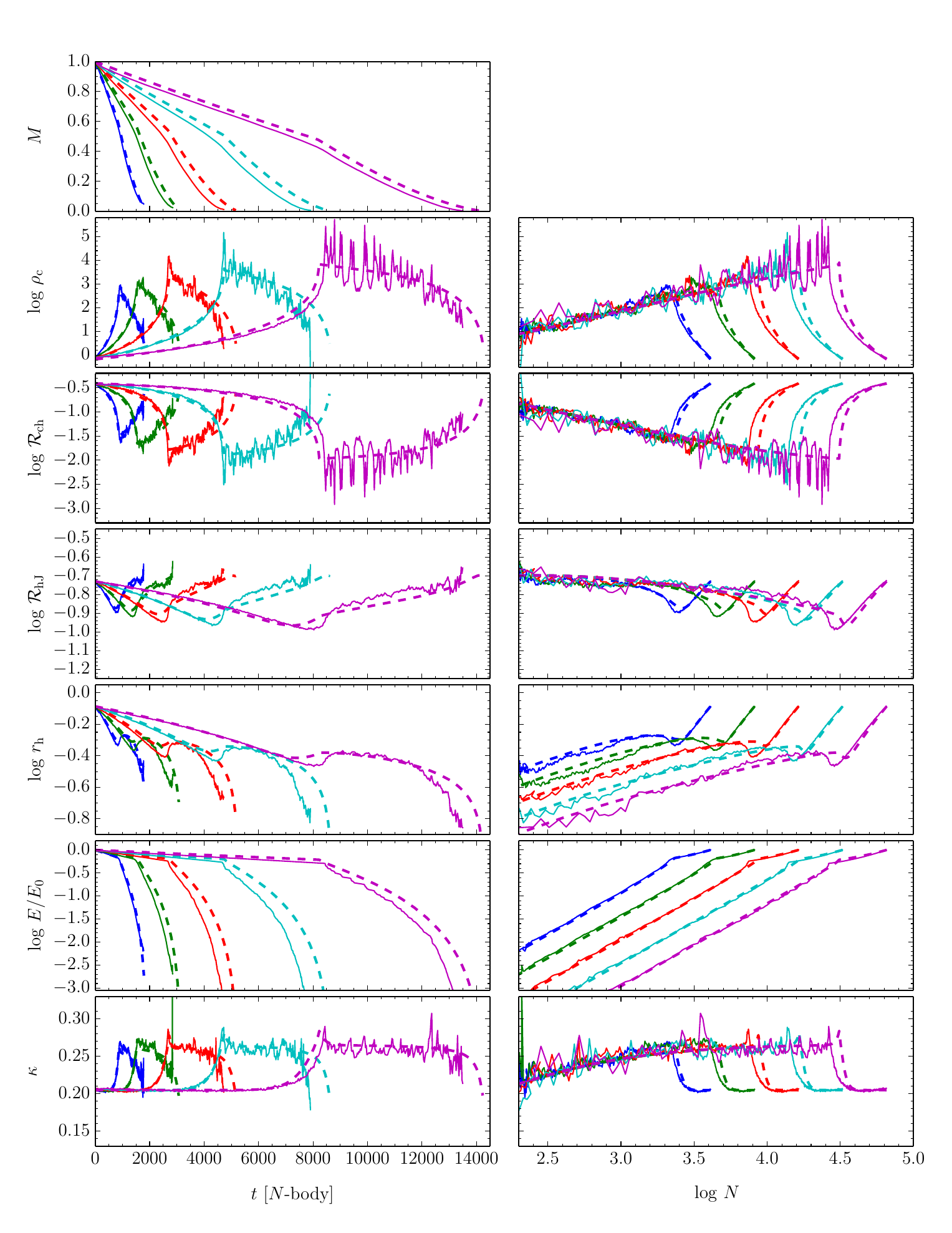}
\caption{Evolution of all cluster parameters as a function of $N$-body time (left) and $N$ (right). In the left panels the evolution is from left to right, and in the right-hand panels evolution is from right to left. In the top panel ($M(t)$), the different curves from left to right are for $N=4\,096$ (blue), $N=8\,192$ (green), $N=16\,384$ (red), $N=32\,768$ and $N=65\,536$ (magenta). The initial $\rhrj\simeq 1/5$ and the initial conditions for $N$-body models were given by a \citet{1966AJ.....71...64K} with $W_0 = 5$. The evolution of all parameters in the unbalanced phase (roughly first half of the evolution) and the balanced phase (roughly second half) is well described by the new version of \emacss\ (shown as dashed lines).}
\label{fig:nbody_rjrh5}
\end{figure*}

\begin{figure*}
\center\includegraphics[width=16cm]{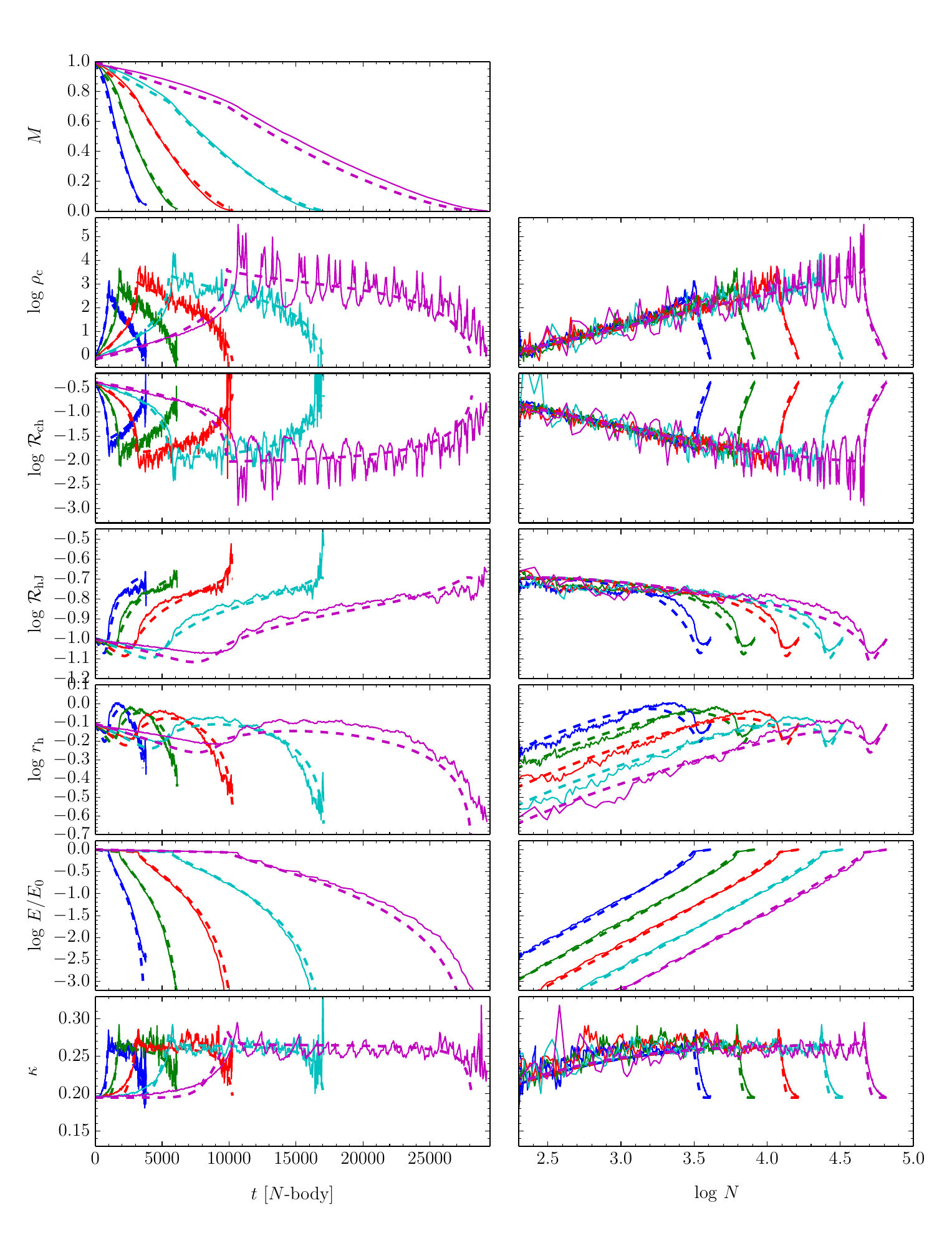}
\caption{As Fig.~\ref{fig:nbody_rjrh5}, but now for $\rhrj(0) = 1/10$ and a Plummer model as initial conditions. In the left panels the evolution is from left to right, and in the right-hand panels evolution is from right to left.}
\label{fig:nbody_rjrh10}
\end{figure*}

In Figs~\ref{fig:nbody_rjrh5} and \ref{fig:nbody_rjrh10} we show the results of the evolution of all parameters for the $N$-body runs with initial $\rhrj = 1/5$ and $1/10$, respectively. The results following from \emacss\ are shown as dashed lines and provide an accurate description of $\rc$, $\rh$, $\rj$ (i.e. $N$). The evolution of the derived quantities $\rhoc$, $E$ and $\kappa$ are also well reproduced. If we consider the temporal aspects of evolution, such as the moment of core collapse and the total lifetime then  the difference between the \emacss\ results and the $N$-body results is within approximately 10\% for these models.

\begin{figure*}
\center\includegraphics[width=16cm]{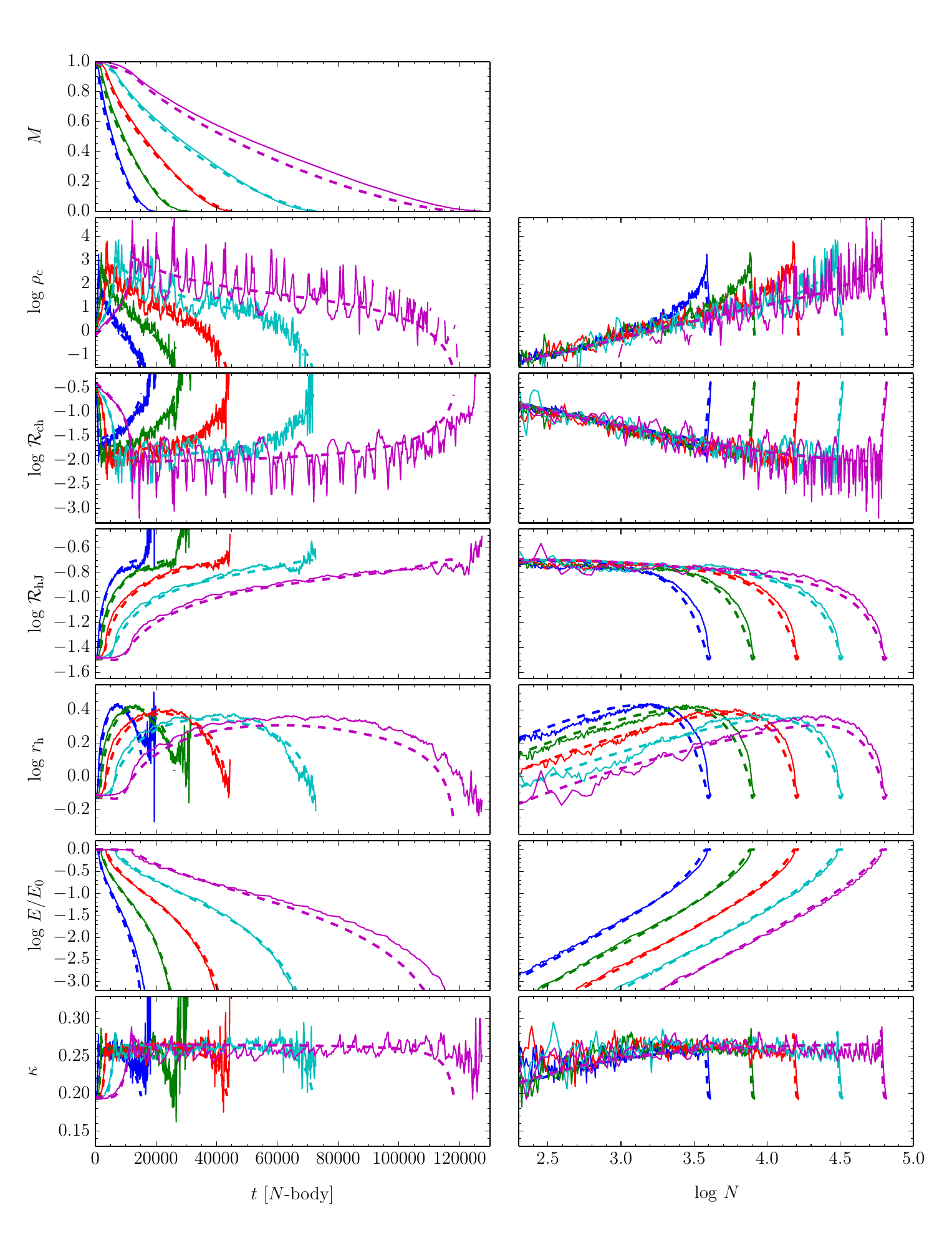}
\caption{As Fig.~\ref{fig:nbody_rjrh5}, but now for $\rhrj = 1/30$ and a Plummer model as initial conditions. In the left panels the evolution is from left to right, and in the right-hand panels evolution is from right to left. These $N$-body models were first presented in Paper~I.}
\label{fig:nbody_rjrh30}
\end{figure*}

\begin{figure*}
\center\includegraphics[width=16cm]{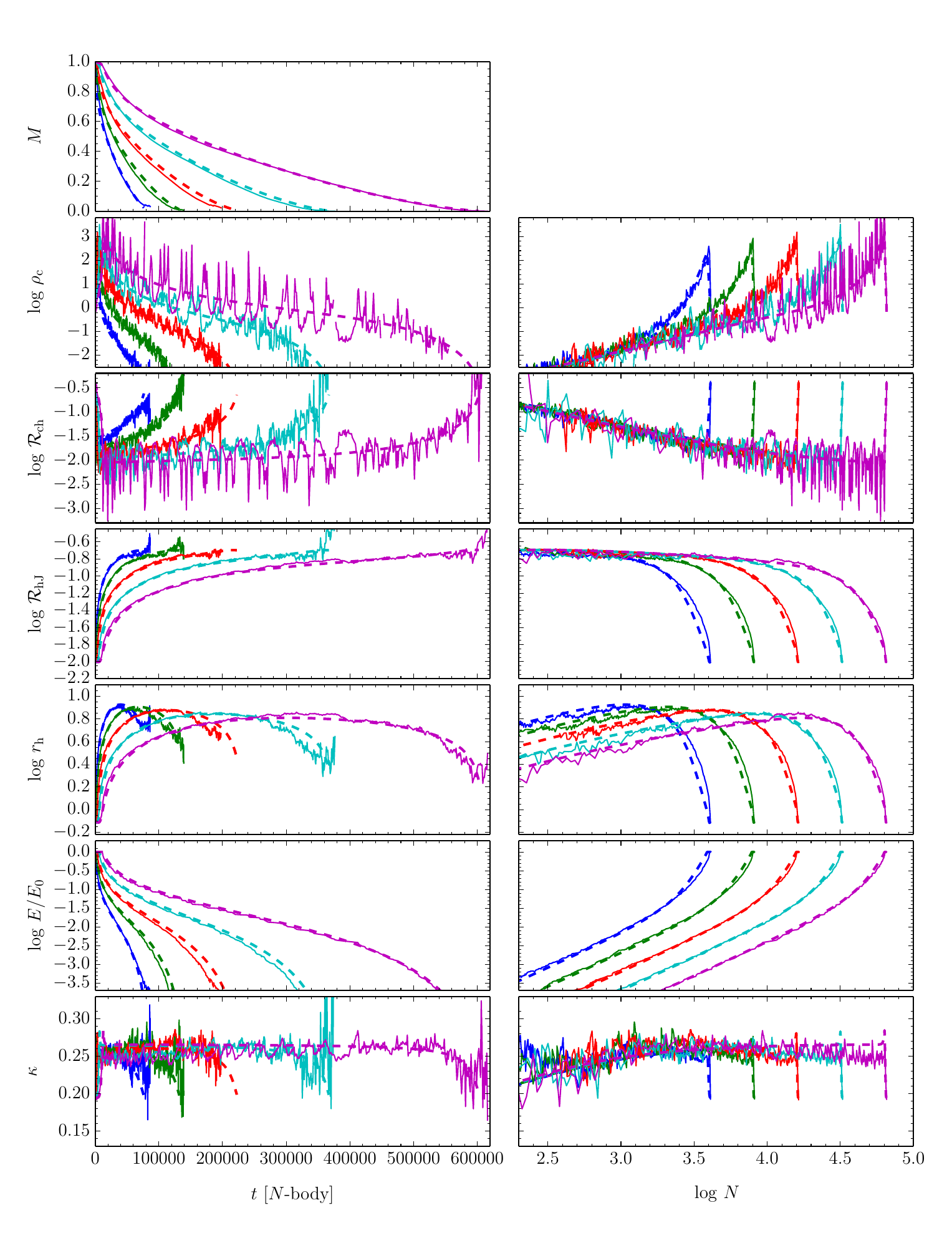}
\caption{As Fig.~\ref{fig:nbody_rjrh5}, but now for $\rhrj = 1/100$ and a Plummer model as initial conditions. In the left panels the evolution is from left to right, and in the right-hand panels evolution is from right to left. These $N$-body models were first presented in Paper~I.}
\label{fig:nbody_rjrh100}
\end{figure*}

Figs~\ref{fig:nbody_rjrh30} and ~\ref{fig:nbody_rjrh100} show the results for the  compact clusters of Paper~I with initial $\rhrj = 1/30$ and $\rhrj=1/100$, respectively. For these clusters, the evolution of \emacss\ is very similar to the version presented in Paper~I and the good agreement between \emacss\ and the $N$-body models is therefore as expected. A small difference with Paper~I is that we here compare the model to the half-mass radius $\rh$ and the virial radius $\rv$ (through $\kappa$), whereas in Paper~I we only considered the virial radius because we assumed $\rh=\rv$. 

The only parameters we have not discussed yet are $N_2$  and $N_3$ (equation~\ref{eq:rcpost}). From a comparison of \emacss\ to the asymptotic evolution of $\rcrh(N)$ we find $N_2=12$. The model is not very sensitive to the exact value of $N_3$. Clusters with $N\gtrsim N_3$ evolve at constant $\rcrh$ in the balanced phase, whereas clusters with $N\lesssim N_3$ evolve as $\rcrh\propto N^{-2/3}$ (Section~\ref{ssec:coreevol}). We find that for a value of $N_3 = N_1 = 15\,000$ \emacss\ provides a  satisfactory  description of $\rc$ for all runs.  A summary of all model parameters is given in Table~\ref{tab:parameters}.

%%%%%%%%%%%%%%%%%%%%%%%%%%%%%%%%%%%%%%%%%%%%
\section{Conclusions and future work}
\label{sec:conclusion}
The new version of \emacss\  reproduces the evolution of the three fundamental  radii of single-mass clusters evolving in a steady tidal field: the core radius $\rc$, the half-mass radius $\rh$ and the Jacobi (or tidal) radius $\rj$, where the latter is equivalent to the evolution of the total mass $M$, or the number of stars $N$. Compared to version one (Paper~I) the code now also reproduces the unbalanced evolution which is important for  clusters in strong tidal fields (i.e. large initial $\rhrj$). This version also introduces the evolution of the core density $\rhoc$ and a related cluster concentration parameter $\kappa= \rh/(4\rv)$ that depends on the ratio $\rcrh=\rc/\rh$. The evolution of the core parameters introduces an additional number of integrations steps compared to Paper~I, most of which are in the phase just before core collapse (the gravothermal catastrophe), when the core contracts on a core relaxation time. Still, the entire evolution is solved with a modest number of about 2000 integration steps,  such that about $10^3$ models can be computed in a second on a single-core desktop computer.

In a follow-up paper (Alexander et al., in preparation, Paper~III), we expand \emacss\ to reproduce clusters with more realistic (initial) properties such as a stellar mass function, and the evolution and mass-loss of  stars. Both code modules (single-mass and multi-mass) will be available in the same code and a command line switch allows the user to select one of them.
 It is worth noting that the computational effort for solving cluster evolution is almost $N$-independent, which makes \emacss\ a powerful tool to do population synthesis studies of globular cluster populations (\citealt{2013MNRAS.432L...1A}; Alexander et al., in preparation).

%%%%%%%%%%%%%%%%%%%%%%%%%%%%%%%%%%%%%%%%%%%%
\section*{Acknowledgement}
PA acknowledges STFC for financial support. MG acknowledges financial support from the Royal Society in the form of a University Research Fellowship (URF) and an equipment grant that was used to purchase nodes equipped  with Graphics Processing Units (GPUs)  that were used for the $N$-body computations. All authors thank the Royal Society for an International Exchange Grant between the UK and the University of Queensland in Brisbane. HB is supported by the Australian Research Council through Future Fellowship grant FT0991052. All authors thank Sverre Aarseth for his support of {\small\textsc{NBODY6}} and Keigo Nitadori for the GPU implementation. Douglas Heggie is acknowledged for several interesting discussions  and for constructive  comments on the manuscript.  The authors thank the referee Mirek Giersz for carefully reading the paper and for providing  constructive comments.
\bibliographystyle{mn2e}

\begin{thebibliography}{}

\bibitem[\protect\citeauthoryear{{Aarseth}}{{Aarseth}}{1999}]{1999PASP..111.1333A}
{Aarseth} S.~J.,  1999, \pasp, 111, 1333

\bibitem[\protect\citeauthoryear{{Aarseth}}{{Aarseth}}{2003}]{2003gnbs.book.....A}
{Aarseth} S.~J.,  2003, {Gravitational N-Body Simulations}.
{Cambridge University Press}

\bibitem[\protect\citeauthoryear{{Aarseth} \& {Heggie}}{{Aarseth} \&
  {Heggie}}{1998}]{1998MNRAS.297..794A}
{Aarseth} S.~J.,  {Heggie} D.~C.,  1998, \mnras, 297, 794

\bibitem[\protect\citeauthoryear{{Aarseth}, {Henon} \& {Wielen}}{{Aarseth}
  et~al.}{1974}]{1974A&A....37..183A}
{Aarseth} S.~J.,  {Henon} M.,    {Wielen} R.,  1974, \aap, 37, 183

\bibitem[\protect\citeauthoryear{{Ahmad} \& {Cohen}}{{Ahmad} \&
  {Cohen}}{1973}]{1973JCoPh..12..389A}
{Ahmad} A.,  {Cohen} L.,  1973, J. of Comput. Phys., 12, 389

\bibitem[\protect\citeauthoryear{{Alexander} \& {Gieles}}{{Alexander} \&
  {Gieles}}{2012}]{2012MNRAS.422.3415A}
{Alexander} P.~E.~R.,  {Gieles} M.,  2012, \mnras, 422, 3415 (Paper~I)

\bibitem[\protect\citeauthoryear{{Alexander} \& {Gieles}}{{Alexander} \&
  {Gieles}}{2013}]{2013MNRAS.432L...1A}
{Alexander} P.~E.~R.,  {Gieles} M.,  2013, \mnras, 432, L1

\bibitem[\protect\citeauthoryear{{Allen} \& {Heggie}}{{Allen} \&
  {Heggie}}{1992}]{1992MNRAS.257..245A}
{Allen} F.~S.,  {Heggie} D.~C.,  1992, \mnras, 257, 245

\bibitem[\protect\citeauthoryear{{Baumgardt}}{{Baumgardt}}{2001}]{2001MNRAS.325.1323B}
{Baumgardt} H.,  2001, \mnras, 325, 1323

\bibitem[\protect\citeauthoryear{{Baumgardt}, {Heggie}, {Hut} \&
  {Makino}}{{Baumgardt} et~al.}{2003}]{2003MNRAS.341..247B}
{Baumgardt} H.,  {Heggie} D.~C.,  {Hut} P.,    {Makino} J.,  2003, \mnras, 341,
  247

\bibitem[\protect\citeauthoryear{{Baumgardt}, {Hut} \& {Heggie}}{{Baumgardt}
  et~al.}{2002}]{2002MNRAS.336.1069B}
{Baumgardt} H.,  {Hut} P.,    {Heggie} D.~C.,  2002, \mnras, 336, 1069

\bibitem[\protect\citeauthoryear{{Bettwieser} \& {Sugimoto}}{{Bettwieser} \&
  {Sugimoto}}{1984}]{1984MNRAS.208..493B}
{Bettwieser} E.,  {Sugimoto} D.,  1984, \mnras, 208, 493

\bibitem[\protect\citeauthoryear{{Brodie} \& {Strader}}{{Brodie} \&
  {Strader}}{2006}]{2006ARA&A..44..193B}
{Brodie} J.~P.,  {Strader} J.,  2006, \araa, 44, 193

\bibitem[\protect\citeauthoryear{{Cohn}}{{Cohn}}{1980}]{1980ApJ...242..765C}
{Cohn} H.,  1980, \apj, 242, 765

\bibitem[\protect\citeauthoryear{{Eldridge}, {Izzard} \& {Tout}}{{Eldridge}
  et~al.}{2008}]{2008MNRAS.384.1109E}
{Eldridge} J.~J.,  {Izzard} R.~G.,    {Tout} C.~A.,  2008, \mnras, 384, 1109

\bibitem[\protect\citeauthoryear{{Freeman} \& {Bland-Hawthorn}}{{Freeman} \&
  {Bland-Hawthorn}}{2002}]{2002ARA&A..40..487F}
{Freeman} K.,  {Bland-Hawthorn} J.,  2002, \araa, 40, 487

\bibitem[\protect\citeauthoryear{{Fukushige} \& {Heggie}}{{Fukushige} \&
  {Heggie}}{1995}]{1995MNRAS.276..206F}
{Fukushige} T.,  {Heggie} D.~C.,  1995, \mnras, 276, 206

\bibitem[\protect\citeauthoryear{{Fukushige} \& {Heggie}}{{Fukushige} \&
  {Heggie}}{2000}]{2000MNRAS.318..753F}
{Fukushige} T.,  {Heggie} D.~C.,  2000, \mnras, 318, 753

\bibitem[\protect\citeauthoryear{{Gaburov}, {Harfst} \& {Portegies
  Zwart}}{{Gaburov} et~al.}{2009}]{2009NewA...14..630G}
{Gaburov} E.,  {Harfst} S.,    {Portegies Zwart} S.,  2009, \na, 14, 630

\bibitem[\protect\citeauthoryear{{Gieles} \& {Baumgardt}}{{Gieles} \&
  {Baumgardt}}{2008}]{2008MNRAS.389L..28G}
{Gieles} M.,  {Baumgardt} H.,  2008, \mnras, 389, L28

\bibitem[\protect\citeauthoryear{{Gieles}, {Baumgardt}, {Heggie} \&
  {Lamers}}{{Gieles} et~al.}{2010}]{2010MNRAS.408L..16G}
{Gieles} M.,  {Baumgardt} H.,  {Heggie} D.~C.,    {Lamers} H.~J.~G.~L.~M.,
  2010, \mnras, 408, L16

\bibitem[\protect\citeauthoryear{{Gieles}, {Heggie} \& {Zhao}}{{Gieles}
  et~al.}{2011}]{2011MNRAS.413.2509G}
{Gieles} M.,  {Heggie} D.~C.,    {Zhao} H.,  2011, \mnras, 413, 2509

\bibitem[\protect\citeauthoryear{{Giersz} \& {Heggie}}{{Giersz} \&
  {Heggie}}{1994}]{1994MNRAS.268..257G}
{Giersz} M.,  {Heggie} D.~C.,  1994, \mnras, 268, 257

\bibitem[\protect\citeauthoryear{{Giersz} \& {Heggie}}{{Giersz} \&
  {Heggie}}{1997}]{1997MNRAS.286..709G}
{Giersz} M.,  {Heggie} D.~C.,  1997, \mnras, 286, 709

\bibitem[\protect\citeauthoryear{{Gnedin}, {Ostriker} \& {Tremaine}}{{Gnedin}
  et~al.}{2013}]{2013arXiv1308.0021G}
{Gnedin} O.~Y.,  {Ostriker} J.~P.,    {Tremaine} S.,  2013, ArXiv:1308.0021

\bibitem[\protect\citeauthoryear{{Goodman}}{{Goodman}}{1987}]{1987ApJ...313..576G}
{Goodman} J.,  1987, \apj, 313, 576

\bibitem[\protect\citeauthoryear{{Harris}, {Harris} \& {Alessi}}{{Harris}
  et~al.}{2013}]{2013ApJ...772...82H}
{Harris} W.~E.,  {Harris} G.~L.~H.,    {Alessi} M.,  2013, \apj, 772, 82

\bibitem[\protect\citeauthoryear{{Heggie} \& {Hut}}{{Heggie} \&
  {Hut}}{2003}]{2003gmbp.book.....H}
{Heggie} D.,  {Hut} P.,  2003, {in Heggie D., Hut P., eds,The Gravitational Million-Body Problem: A
  Multidisciplinary Approach to Star Cluster Dynamics}.
Cambridge University Press, Cambridge, 372 pp.

\bibitem[\protect\citeauthoryear{{Heggie}}{{Heggie}}{1975}]{1975MNRAS.173..729H}
{Heggie} D.~C.,  1975, \mnras, 173, 729

\bibitem[\protect\citeauthoryear{{Heggie} \& {Mathieu}}{{Heggie} \&
  {Mathieu}}{1986}]{1986LNP...267..233H}
{Heggie} D.~C.,  {Mathieu} R.~D.,  1986, in Hut P., McMillan S., eds, Lecture
  Notes in Physics, Vol. 267, The Use of Supercomputers in Stellar Dynamics,
  Springer-Verlag, Berlin, p.233

\bibitem[\protect\citeauthoryear{{Heggie} \& {Stevenson}}{{Heggie} \&
  {Stevenson}}{1988}]{1988MNRAS.230..223H}
{Heggie} D.~C.,  {Stevenson} D.,  1988, \mnras, 230, 223

\bibitem[\protect\citeauthoryear{{H{\'e}non}}{{H{\'e}non}}{1961}]{1961AnAp...24..369H}
{H{\'e}non} M.,  1961, Annales d'Astrophysique, 24, 369; English translation:
  ArXiv:1103.3499

\bibitem[\protect\citeauthoryear{{H{\'e}non}}{{H{\'e}non}}{1965}]{1965AnAp...28...62H}
{H{\'e}non} M.,  1965, Annales d'Astrophysique, 28, 62; English translation:
  ArXiv:1103.3498

\bibitem[\protect\citeauthoryear{{Hurley}, {Pols} \& {Tout}}{{Hurley}
  et~al.}{2000}]{2000MNRAS.315..543H}
{Hurley} J.~R.,  {Pols} O.~R.,    {Tout} C.~A.,  2000, \mnras, 315, 543

\bibitem[\protect\citeauthoryear{{Hurley} \& {Shara}}{{Hurley} \&
  {Shara}}{2012}]{2012MNRAS.425.2872H}
{Hurley} J.~R.,  {Shara} M.~M.,  2012, \mnras, 425, 2872

\bibitem[\protect\citeauthoryear{{Hurley}, {Tout} \& {Pols}}{{Hurley}
  et~al.}{2002}]{2002MNRAS.329..897H}
{Hurley} J.~R.,  {Tout} C.~A.,    {Pols} O.~R.,  2002, \mnras, 329, 897

\bibitem[\protect\citeauthoryear{{Inagaki} \& {Lynden-Bell}}{{Inagaki} \&
  {Lynden-Bell}}{1983}]{1983MNRAS.205..913I}
{Inagaki} S.,  {Lynden-Bell} D.,  1983, \mnras, 205, 913

\bibitem[\protect\citeauthoryear{{Jord{\'a}n}, {C{\^o}t{\'e}}, {Blakeslee},
  {Ferrarese}, {McLaughlin}, {Mei}, {Peng}, {Tonry}, {Merritt},
  {Milosavljevi{\'c}}, {Sarazin}, {Sivakoff} \& {West}}{{Jord{\'a}n}
  et~al.}{2005}]{2005ApJ...634.1002J}
{Jord{\'a}n} A.,  et al.,
   2005, \apj, 634, 1002

\bibitem[\protect\citeauthoryear{{Jord{\'a}n}, {McLaughlin}, {C{\^o}t{\'e}},
  {Ferrarese}, {Peng}, {Mei}, {Villegas}, {Merritt}, {Tonry} \&
  {West}}{{Jord{\'a}n} et~al.}{2007}]{2007ApJS..171..101J}
{Jord{\'a}n} A.,  et~al.,  2007, \apjs, 171, 101

\bibitem[\protect\citeauthoryear{{Kim}, {Lee} \& {Goodman}}{{Kim}
  et~al.}{1998}]{1998ApJ...495..786K}
{Kim} S.~S.,  {Lee} H.~M.,    {Goodman} J.,  1998, \apj, 495, 786

\bibitem[\protect\citeauthoryear{{King}}{{King}}{1966}]{1966AJ.....71...64K}
{King} I.~R.,  1966, \aj, 71, 64

\bibitem[\protect\citeauthoryear{{Lamers}, {Baumgardt} \& {Gieles}}{{Lamers}
  et~al.}{2010}]{2010MNRAS.409..305L}
{Lamers} H.~J.~G.~L.~M.,  {Baumgardt} H.,    {Gieles} M.,  2010, \mnras, 409,
  305

\bibitem[\protect\citeauthoryear{{Larson}}{{Larson}}{1970}]{1970MNRAS.150...93L}
{Larson} R.~B.,  1970, \mnras, 150, 93

\bibitem[\protect\citeauthoryear{{Lombardi} Jr., {Warren}, {Rasio}, {Sills} \&
  {Warren}}{{Lombardi} et~al.}{2002}]{2002ApJ...568..939L}
{Lombardi} Jr. J.~C.,  {Warren} J.~S.,  {Rasio} F.~A.,  {Sills} A.,    {Warren}
  A.~R.,  2002, \apj, 568, 939

\bibitem[\protect\citeauthoryear{{Lynden-Bell} \& {Eggleton}}{{Lynden-Bell} \&
  {Eggleton}}{1980}]{1980MNRAS.191..483L}
{Lynden-Bell} D.,  {Eggleton} P.~P.,  1980, \mnras, 191, 483

\bibitem[\protect\citeauthoryear{{Lynden-Bell} \& {Wood}}{{Lynden-Bell} \&
  {Wood}}{1968}]{1968MNRAS.138..495L}
{Lynden-Bell} D.,  {Wood} R.,  1968, \mnras, 138, 495

\bibitem[\protect\citeauthoryear{{Makino} \& {Aarseth}}{{Makino} \&
  {Aarseth}}{1992}]{1992PASJ...44..141M}
{Makino} J.,  {Aarseth} S.~J.,  1992, \pasj, 44, 141

\bibitem[\protect\citeauthoryear{{Meylan} \& {Heggie}}{{Meylan} \&
  {Heggie}}{1997}]{1997A&ARv...8....1M}
{Meylan} G.,  {Heggie} D.~C.,  1997, \aapr, 8, 1

\bibitem[\protect\citeauthoryear{{Nitadori} \& {Aarseth}}{{Nitadori} \&
  {Aarseth}}{2012}]{2012MNRAS.424..545N}
{Nitadori} K.,  {Aarseth} S.~J.,  2012, \mnras, 424, 545

\bibitem[\protect\citeauthoryear{{Plummer}}{{Plummer}}{1911}]{1911MNRAS..71..460P}
{Plummer} H.~C.,  1911, \mnras, 71, 460

\bibitem[\protect\citeauthoryear{{Portegies Zwart}, {McMillan} \&
  {Gieles}}{{Portegies Zwart} et~al.}{2010}]{2010ARA&A..48..431P}
{Portegies Zwart} S.~F.,  {McMillan} S.~L.~W.,    {Gieles} M.,  2010, \araa,
  48, 431

\bibitem[\protect\citeauthoryear{{Prieto} \& {Gnedin}}{{Prieto} \&
  {Gnedin}}{2008}]{2008ApJ...689..919P}
{Prieto} J.~L.,  {Gnedin} O.~Y.,  2008, \apj, 689, 919

\bibitem[\protect\citeauthoryear{{Sippel} \& {Hurley}}{{Sippel} \&
  {Hurley}}{2013}]{2013MNRAS.430L..30S}
{Sippel} A.~C.,  {Hurley} J.~R.,  2013, \mnras, 430, L30

\bibitem[\protect\citeauthoryear{{Spitzer}}{{Spitzer}}{1987}]{1987degc.book.....S}
{Spitzer} L.,  1987, {Dynamical Evolution of Globular Clusters}.
Princeton University Press, Princeton, 191 pp.

\bibitem[\protect\citeauthoryear{{Spitzer} \& {Hart}}{{Spitzer} \&
  {Hart}}{1971}]{1971ApJ...164..399S}
{Spitzer} L.~J.,  {Hart} M.~H.,  1971, \apj, 164, 399

\end{thebibliography}

\end{document}